\documentclass[twocolumn,british,prl,superscriptaddress]{revtex4-1}
\usepackage[T1]{fontenc}
\usepackage[latin9]{inputenc}
\usepackage{amsmath}
\usepackage{amssymb}
\usepackage{graphicx}
\usepackage{babel}
\usepackage{xcolor}
\begin{document}
\title{Wrinkle patterns in active viscoelastic thin sheets}
\author{D. A. Matoz-Fernandez}
\email{d.a.matozfernandez@dundee.ac.uk}
\affiliation{School of Life Sciences, University of Dundee, 
Dundee, UK DD1 5EH}
\author{Fordyce A. Davidson}
\affiliation{School of Science and Engineering, University of Dundee, 
Dundee, UK DD1 5EN}
\author{Nicola R. Stanley-Wall}
\affiliation{School of Life Sciences, University of Dundee, 
Dundee, UK DD1 5EH}
\author{Rastko Sknepnek}
\email{r.sknepnek@dundee.ac.uk}
\affiliation{School of Life Sciences, University of Dundee, Dundee, UK DD1 5EH}
\affiliation{School of Science and Engineering, University of Dundee, Dundee, UK DD1 5EN}

\begin{abstract}
We show that a viscoelastic thin sheet driven out of equilibrium by
active structural remodelling develops a rich
variety of shapes as a result of a competition between viscous relaxation
and activity. In the regime where active processes are faster than
viscoelastic relaxation, wrinkles that are formed due to remodelling
are unable to relax to a configuration that minimises the elastic
energy and the sheet is inherently out of equilibrium. We argue that
this non-equilibrium regime is of particular interest in biology as
it allows the system to access morphologies that are unavailable if
restricted to the adiabatic evolution between configurations that
minimise the elastic energy alone. Here, we introduce activity using
the formalism of evolving target metric and showcase the diversity
of wrinkling morphologies arising from out of equilibrium dynamics.
\end{abstract}
\maketitle
D'Arcy Thompson set the mathematical foundation for describing and
classifying the astonishing diversity of shapes and form in the living
world \citep{thompson1942growth}. A century later, our understanding
of biological processes at the molecular level has been vastly improved
\citep{Alberts2002}, yet it is still largely unknown how the formation
of large, functional structures such as tissues and organs arises
from these molecular processes \citep{wolpert2015principles}. A unifying
feature of all higher organisms is that they start as a single cell,
a zygote, and autonomously develop into an individual, without external
input. The genome provides a template that steers development towards
the desired body plan \citep{wolpert2015principles}. The formation
of large structures such as tissues and organs is a result of a complex
set of guided collective mechano-chemical processes. To select a specific
morphology, the phase space of possible shapes has to be large. Furthermore,
transition between shapes should be possible at a reasonably low cost,
which is hard to achieve in equilibrium.

Out of equilibrium biological processes are naturally described within
the framework of the active matter physics, where the system is driven
out of equilibrium by a constant input of energy at the microscopic
scale \citep{marchetti2013hydrodynamics}. Despite great progress
in understanding the behaviour of active fluids, much less is known
about how activity affects the behaviour of solid and viscoelastic
materials, such as tissues \citep{harris2012characterizing,berthoumieux2014active,matoz2017nonlinear}.
Numerical simulations of dense self-propelled elastic disks, for example,
showed that part of the energy intake is diverted into local elastic
deformations leading to prominent spatial and temporal heterogeneities
in observed velocity fields \citep{Henkes2011}. Such dynamical heterogeneity
is a hallmark of an active glassy state \citep{berthier2013non},
with epithelial cell monolayers being prime examples of such behaviour
\citep{petitjean2010velocity,angelini2011glass,chepizhko2018jamming,henkes2019universal}.
The biological significance of dynamical heterogeneity is only starting
to emerge. When it comes to describing bending deformations in active
systems, only recently a theoretical description has been proposed
\citep{salbreux2017mechanics}.

\begin{figure}
\begin{centering}
\includegraphics[width=0.95\columnwidth]{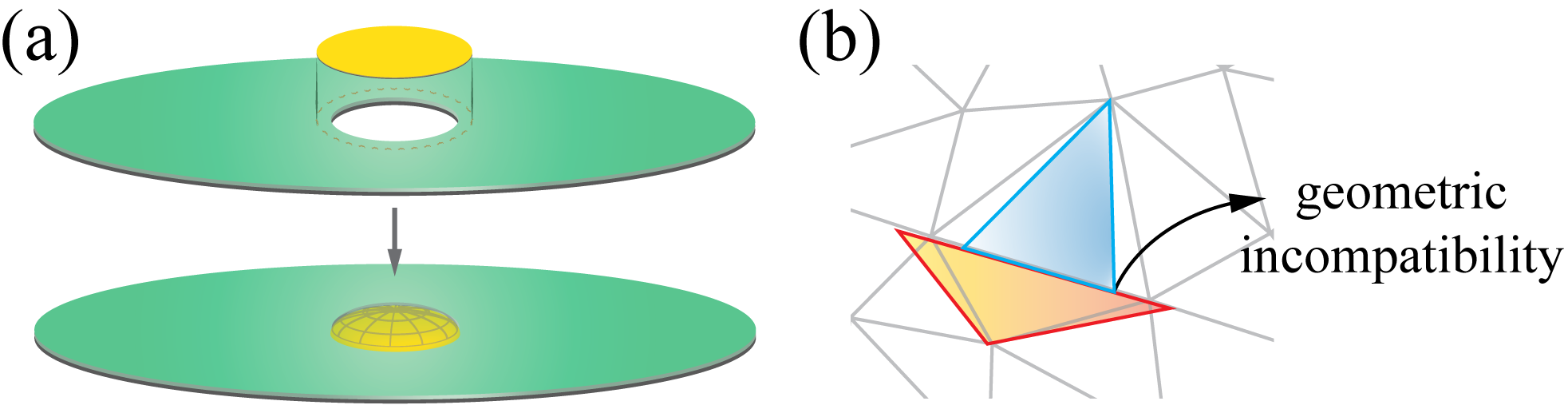}
\par\end{centering}
\caption{(a) Inserting an elastic disk into an aperture smaller than disk's
size (top) induces residual stress in the disk due to compression.
This residual stress can be released by buckling out of plane (bottom).
(b) In a discrete picture where an elastic sheet is represented as
a triangulation of a surface, the geometric incompatibility leading
to the residual stress (i.e., non-embeddable metric) can be understood
as two triangles that share an edge having mutually incompatible preferred
shapes, e.g. red (blue) triangle is the preferred shape for the corresponding
grey mesh triangle shown underneath it. \label{fig:embed}}
\end{figure}

In this paper, we study thin elastic and viscoelastic sheets with
activity introduced as a dynamical change of the reference shape.
Physically, activity provides structural remodelling that acts as
a local time-dependent source of strain. The time-dependent reference
shape can be either stress free (embeddable metric) or contain residual
stress (non-embeddable metric) \citep{efrati2009elastic}. While the
distinction between the two cases has important consequences for the
elastic ground state \citep{efrati2009elastic,kang2014complex}, it
is not essential for the present discussion. As shown in Fig.~\ref{fig:embed},
bending out of plane can fully or partly remove the residual stresses
due to remodelling, depending on whether the particular reference
state is embeddable or not in $\mathbb{R}^{3}$. It has been recently
argued \citep{clement2017viscoelastic} that viscoelastic relaxation
can stabilise cell shapes during morphogenesis. Such viscoelastic
effects remove all stresses over a sufficiently long time. Here, we
focus on the regime where active remodelling is faster than both elastic
and viscoelastic relaxation, leading to the system being inherently
out of equilibrium. This regime is expected to be of particular importance
to early embryonic development.

\begin{figure}
\begin{centering}
\includegraphics[width=0.95\columnwidth]{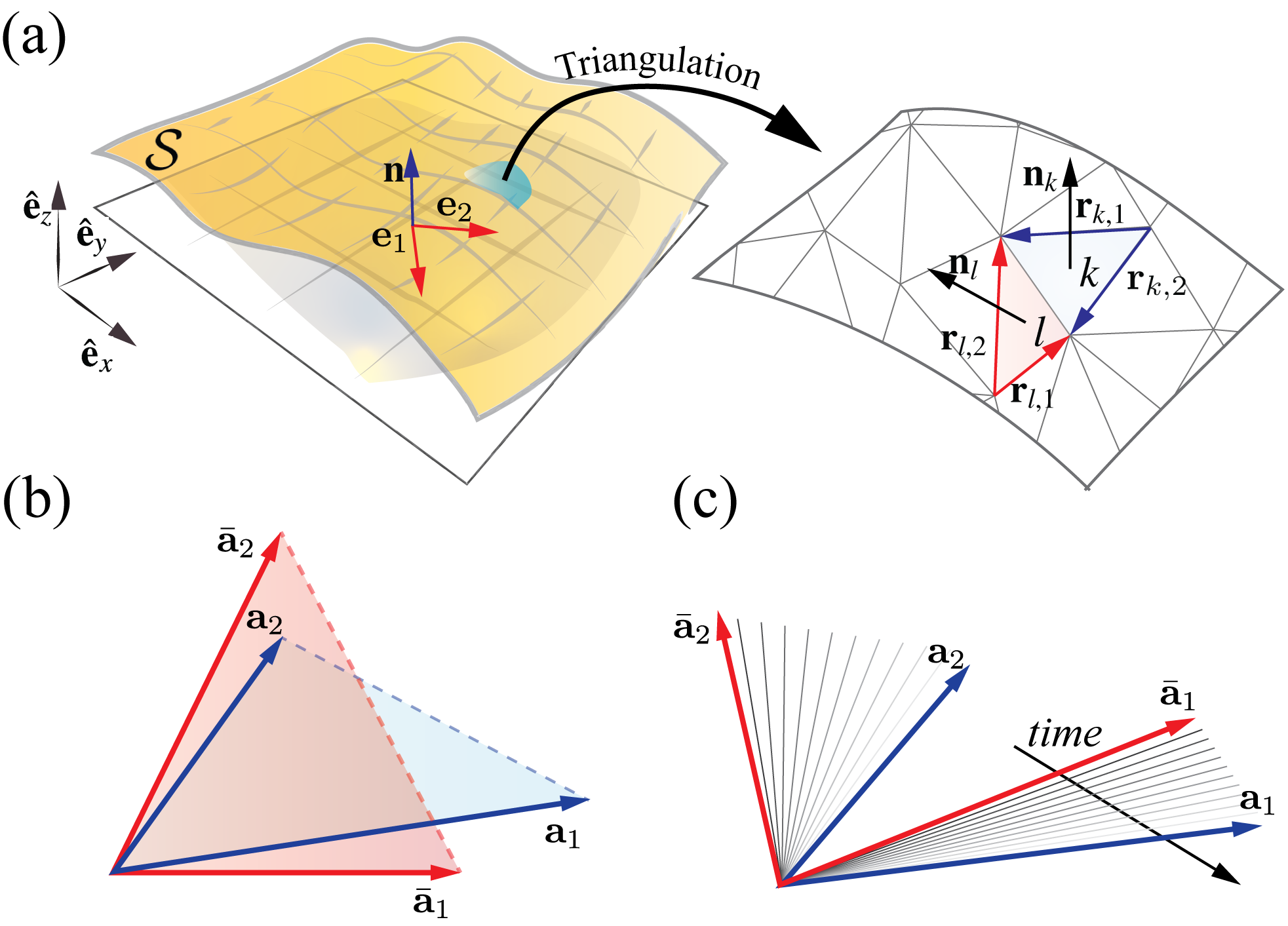}
\par\end{centering}
\caption{(a) The sheet is represented as a two-dimensional mid-surface parametrised
by $\left(x,y\right)-$coordinates, with two tangent vectors (\textbf{$\mathbf{e}_{1}\equiv\partial_{x}\mathbf{r}$},
$\mathbf{e}_{2}\equiv\partial_{y}\mathbf{r}$) and a unit-length normal
$\left(\mathbf{n}=\left(\mathbf{e}_{1}\times\mathbf{e}_{2}\right)/\left|\mathbf{e}_{1}\times\mathbf{e}_{2}\right|\right)$
assigned to each point of the surface. For numerical implementation,
the surface is discretised in terms of triangles. (b) Red vectors
form the reference metric tensor, $\bar{g}_{\alpha\beta}=\bar{\mathbf{a}}_{\alpha}\cdot\bar{\mathbf{a}}_{\beta}$
and blue vectors form the realised metric tensor, $g_{\alpha\beta}=\mathbf{a}_{\alpha}\cdot\mathbf{a}_{\beta}$.
The strain tensor is defined as $u_{\alpha\beta}=\frac{1}{2}\left(g_{\alpha\beta}-\overline{g}_{\alpha\beta}\right)$.
(c) Viscoelasticity is modelled as a relaxation of the reference metric
towards the realised metric, with a characteristic time scale $\tau_{v}$.
\label{fig:model}}
\end{figure}

We study a thin sheet of size $L$ and uniform thickness $h\ll L$
with linear elastic response \citep{Audoly2010}. We assume that the
surrounding fluid provides damping but ignore all other hydrodynamic
effects. The sheet is represented by the two-dimensional mid-surface,
initially in the $xy$ plane. The deformed mid-surface with no overhangs
can be parametrised as $\mathbf{r}=\mathbf{r}\left(x,y\right)=\left(x,y,w\left(x,y\right)\right)$,
where $w\left(x,y\right)$ is a sufficiently smooth height function.
One defines the metric, $g_{\alpha\beta}=\partial_{\alpha}\mathbf{r}\cdot\partial_{\beta}\mathbf{r}$,
and curvature, $c_{\alpha}^{\beta}=g^{\beta\gamma}b_{\alpha\gamma}$,
tensors where $b_{\alpha\beta}=-\partial_{\alpha}\mathbf{r}\cdot\partial_{\beta}\mathbf{n}$
($\alpha,\beta\in\left\{ x,y\right\} $) is the second fundamental
form and $\mathbf{n}=\left(\partial_{x}\mathbf{r}\times\partial_{y}\mathbf{r}\right)/\left|\partial_{x}\mathbf{r}\times\partial_{y}\mathbf{r}\right|$
is the unit normal vector \citep{do1976differential} (Fig.~\ref{fig:model}a).
The elastic energy of the mid-surface is \citep{koiter1966nonlinear,efrati2009elastic,SI}
\begin{equation}
E=\int dA\mathcal{A}^{\alpha\beta\gamma\delta}\left(\frac{h}{2}u_{\alpha\beta}u_{\gamma\delta}+\frac{h^{3}}{24}b_{\alpha\beta}b_{\gamma\delta}\right),\label{eq:elastic_elergy}
\end{equation}
where $u_{\alpha\beta}=\frac{1}{2}\left(g_{\alpha\beta}-\overline{g}_{\alpha\beta}\right)$
is the strain tensor, $\overline{g}_{\alpha\beta}$ is a reference
metric tensor, $dA=\sqrt{\det g}dxdy$ is the area element, $\mathcal{A}^{\alpha\beta\gamma\delta}$
is the elastic tensor, and summation over pairs of repeated indices
is assumed. Latin indices refer to the components of vectors in the
embedding Euclidean $\mathbb{R}^{3}$ space, while Greek indices are
used to label intrinsic curvilinear coordinates. For an isotropic
material, $\mathcal{A}^{\alpha\beta\gamma\delta}=\frac{Y}{1+\nu}\left(\frac{\nu}{1-\nu}\overline{g}^{\alpha\beta}\overline{g}^{\gamma\delta}+\overline{g}^{\alpha\gamma}\overline{g}^{\beta\delta}\right)$,
where $Y$ is the Young's modulus and $\nu$ is the Poisson ratio
and $\overline{g}^{\alpha\gamma}\overline{g}_{\gamma\beta}=\delta_{\beta}^{\alpha}$.
The first term in Eq.~(\ref{eq:elastic_elergy}) is the stretching
energy and the second term accounts for bending. For an isotropic
material, stretching and bending energies simplify to $E_{s}=\frac{h}{2}\int dA\frac{Y}{1+\nu}\left(\frac{\nu}{1-\nu}u_{\alpha}^{\alpha}u_{\beta}^{\beta}+u_{\alpha}^{\beta}u_{\beta}^{\alpha}\right)$
and $E_{b}=\frac{h^{3}}{24}\int dA\frac{Y}{1+\nu}\left(\frac{\nu}{1-\nu}c_{\alpha}^{\alpha}c_{\beta}^{\beta}+c_{\alpha}^{\beta}c_{\beta}^{\alpha}\right)$,
with $u_{\alpha}^{\beta}=\overline{g}^{\beta\gamma}u_{\alpha\gamma}$
and $c_{\alpha}^{\beta}=\overline{g}^{\beta\gamma}b_{\alpha\gamma}$
\citep{efrati2009elastic,SI}. With the mean curvature $H=\frac{1}{2}c_{\alpha}^{\alpha}\equiv\frac{1}{2}\text{Tr}\left(\hat{c}\right)$
and the Gaussian curvature $K=\det\left(c_{\alpha}^{\beta}\right)$,
the bending energy becomes $E_{b}=\int dA\kappa\left(2H^{2}-\left(1-\nu\right)K\right)$,
where $\kappa=h^{3}Y/12\left(1-\nu^{2}\right)$ is the bending stiffness.
In general, material properties and the reference metric can be position
dependent and the sheet can have a spontaneous curvature, $H_{0}$.
Here we assume that $H_{0}=0$ and the active remodelling does not
affect elastic parameters. In reality, material properties are affected
by the structural remodelling. However, imposing spatial and time
dependence on the elastic parameters did not qualitatively change
our findings and, for simplicity, in following we assume them be constant.
Finally, we estimate that the relaxation time associated with bending,
$\tau_{el}\sim\eta L^{3}/\kappa$, here $\eta$ is the dynamical viscosity
of the surrounding fluid \citep{SI}. For an epithelial cell sheet
in water, $\tau_{el}\sim10^{1}-10^{2}$ s, consistent with \citep{marmottant2009role}.
Clearly, the time scale of relaxation associated with stretching deformation
is much shorter and consequently of no importance for the present
discussion.

Active effects in a tissue result, for example, from myosin driven
contractions and turnover of the actin cytoskeleton \citep{joanny2009active}
as well as cell growth and division. Processes related to the cytoskeleton
typically occur at time scales of $\tau_{a}\sim10^{1}-10^{2}$ s \citep{rauzi2008nature,rauzi2010planar},
while cell growth and division are slower and can span several hours
\citep{Alberts2002}. Dissipation in tissues results from multi-cellular
rearrangements (i.e., plastic events such as intercalations, ingressions
and extrusions) and sub-cellular cytoskeleton remodelling (i.e., cell
shape relaxation). We note that dissipation is accompanied by entropy
production and, in general, an entropy production equation would be
required \citep{salbreux2017mechanics}. Here, we are not concerned
by the details of the dissipative processes (rendering the entropy
production equation unnecessary) and assume that they occur on a time
scale, $\tau_{v}$. We note, however, that cell rearrangements are
typically slower (occurring on the scale $\sim10$ min) than the sub-cellular
remodelling (occurring on the seconds to minutes scale). While it
is not always the case, the out of equilibrium situation with $\tau_{a}<\tau_{el},\tau_{v}$
is, therefore, biologically plausible and, we argue, beneficial to
access diversity of shapes needed to form complex structures. In the
following, we explore the range of possible dynamical shape patterns
formed in the non-equilibrium regime.

The advantage of expressing deformation with respect to the reference
metric \citep{sknepnek2012nonlinear} is that the formalism can be
directly generalised to include active remodelling and viscoelastic
relaxation, without making only assumptions about the existence of
a stress free reference state. Here, active remodelling is introduced
by imposing dynamical changes of the reference metric. The precise
functional form of active remodelling is not important, as long as
one can associate a typical time scale, $\tau_{a}$, to it. Active
remodelling can be thought of as a generalisation of growth, with
the quasi-static differential growth being described as $\bar{g}_{\alpha\beta}\left(\mathbf{r},t\right)=a\left(\mathbf{r}\right)t\bar{g}_{\alpha\beta}\left(\mathbf{r},t=0\right)$,
where $a\left(\mathbf{r}\right)>0$ and $\tau_{a}^{growth}\equiv a^{-1}\ll\tau_{el}$.
We model viscoelastic effects as a relaxation of the reference metric
towards the realised metric (Fig.~\ref{fig:model}c). Therefore,
viscoelastic relaxation has the opposite effect of elasticity, for
which the reference state conforms to the realised shape rather than
the other way around. A description based on the time-evolving reference
metric is also suitable for direct discretisation (Fig.~\ref{fig:model})
and efficient parallel implementation on GPUs \citep{SI}. This allows
us to simulate systems containing up to $2\times10^{6}$ triangles
removing the need to implement complex remeshing procedures to avoid
reduction in accuracy in the vicinity of high-curvature folds.

We assume overdamped dynamics and solve the set of first-order equations
for each vertex $i$ and discrete metric of each triangle,\begin{subequations}
\begin{align}
\gamma\dot{\mathbf{r}}_{i} & =-\nabla_{\mathbf{r}_{i}}E\left(g_{\alpha\beta},\bar{g}_{\alpha\beta},b_{\alpha\beta}\right)+\boldsymbol{\eta}_{i}\left(t\right),\label{eq:eq_pos}\\
\dot{\bar{g}}_{\alpha\beta}\left(\mathbf{r},t\right) & =R_{\alpha\beta}\left(\overline{g}_{\alpha\beta},t\right)+V_{\alpha\beta}^{\gamma\delta}\left(t\right)\left(g_{\gamma\delta}\left(\mathbf{r},t\right)-\bar{g}_{\gamma\delta}\left(\mathbf{r},t\right)\right).\label{eq:eq_metric}
\end{align}
\end{subequations} Here $\mathbf{r}_{i}\in\mathbb{R}^{3}$ is the
position vector of vertex $i$, $\boldsymbol{\eta}_{i}\left(t\right)\in\mathbb{R}^{3}$
is a weak random noise, obeying $\left\langle \boldsymbol{\eta}_{i}\right\rangle =0$
and $\left\langle \eta_{i}^{m}\left(t\right)\eta_{j}^{n}\left(t^{\prime}\right)\right\rangle =\sqrt{2\gamma k_{B}T}\delta_{ij}\delta_{mn}\delta\left(t-t^{\prime}\right)$
with $m,n\in\left\{ x,y,z\right\} $. $\gamma$ is the friction coefficient
modelling dissipation by the surrounding fluid and $T$ is the temperature
kept very low and used only for numerical convenience to avoid being
trapped in shallow local minima. All our simulations were effectively
at $T=0$ as thermal fluctuations are not expected to play an appreciable
role in biological systems, i.e., relevant energy scales far exceed
$k_{B}T$. $V_{\alpha\beta}^{\gamma\delta}$$\left(t\right)$ is a
tensor that sets the rate of viscous relaxation. While in general
$V_{\alpha\beta}^{\gamma\delta}$ is a function of time, here we assume
it to be constant, $V_{\alpha\beta}^{\gamma\delta}=\frac{1}{\tau_{v}}\delta_{\alpha}^{\gamma}\delta_{\beta}^{\delta}$.
$R_{\alpha\beta}$ is a tensor function that prescribes active remodelling
rate. Here, $R_{\alpha\beta}$ models metric expansion and is given
in Eq. (S34) in \citep{SI}. Furthermore, $R_{\alpha\beta}$ explicitly
depends on time and, thus, models dynamical changes of the active
remodelling rate. Finally, discrete versions of the realised and reference
metric tensors are defined in Fig.~\ref{fig:model}b. Eqs.~(\ref{eq:eq_pos})
and (\ref{eq:eq_metric}) are integrated numerically using standard
first-order Euler-Maruyama discretisation scheme keeping connectivity
of the triangulation fixed. Expressions for the gradient of energy
in Eq.~(\ref{eq:eq_pos}) are straightforward but lengthy \citep{SI}.
Note that in the current implementation, we do not include steric
effects and the sheet can take unphysical self-intersecting configurations.
Including self-avoidance is possible but technically challenging to
efficiently implement on GPUs. Steric effect would indeed affect the
folding patterns but would not change our main conclusions. Values
of parameters used in simulations are given in \citep{SI}. Moreover,
length is measured in units of $h$, time in units of $t^{*}=\gamma/Yh$
and energy in units of $\kappa$.

\begin{figure}
\begin{centering}
\includegraphics[width=0.95\columnwidth]{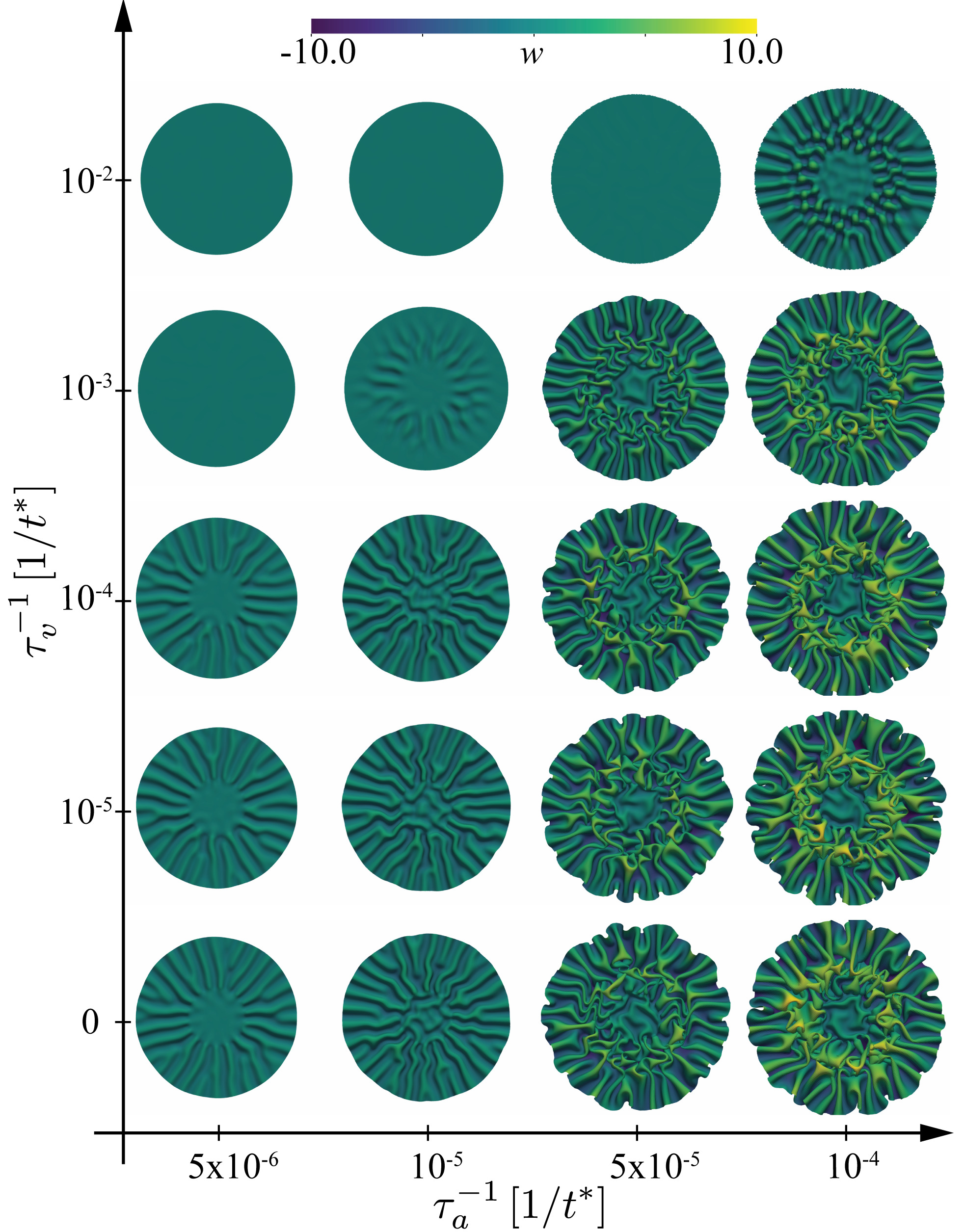}
\par\end{centering}
\caption{A snapshot of the out-of equilibrium shapes obtained by numerical
integration of Eqs.~(\ref{eq:eq_pos}) and (\ref{eq:eq_metric})
starting from a flat disk configuration. The snapshots are taken at
$t=10^{4}t^{*}$. Vertical axis represents the rate of viscous (dissipative)
relaxation with increasing values designating faster residual stress
relaxation. On the horizontal axis we plot the active structural remodelling
rate, with larger values corresponding to faster changes of the local
reference metric. The usual slow, quasi-equilibrium elastic growth
would correspond to the lower left corner in this graph. Colours represent
the height function, $w\left(x,y\right)$. In these simulations, $R_{\alpha\beta}$
is time independent. \label{fig:results}}
\end{figure}

We explored out of equilibrium dynamics of flat disks of radius $R$
subject to active remodelling and viscous dissipation (Fig.~\ref{fig:results}).
The choice of the disk geometry is inspired by extensive work on wrinkling
patterns due to tension \citep{jagla2007modeling,davidovitch2011prototypical}
or resulting from a quasi-equilibrium growth, e.g., during biofilm
formation \citep{amar2014patterns,yan2019mechanical}. This regime
corresponds to $\tau_{el}\ll\tau_{a}$. We assume that a ring of radius
$r_{i}<R$ is kept fixed but can transmit stress. Active remodelling
is assumed to occur only in the outer annulus, for $r_{i}<r<R$. With
no viscoelastic relaxation and slow active remodelling (lower left
corner in Fig.~\ref{fig:results}), the system is in the extensively
studied quasi-equilibrium differential growth regime. Free expansion
of the outer boundary can relieve part of the stress produced by growth.
There is, however, no such stress relief mechanism in the tangential
direction and the sheet forms a regular pattern of radial wrinkles.
The inner disk, on the other hand, is compressed in both directions
leading to wrinkles with no preferred orientation. If one instead
allows for viscoelastic relaxation while keeping the active remodelling
slow (left column in Fig.~\ref{fig:results}), wrinkles are less
pronounced or, in the case of very fast dissipative relaxation, do
not form at all (top left in Fig.~\ref{fig:results}). This is easy
to understand, as in this regime the stress generated by active remodelling
is dissipated by a fast relaxation of the reference metric of the
sheet. As one increases the remodelling rate (second and third columns
in Fig.~\ref{fig:results}), wrinkling patterns become more pronounced
and less regular, especially close to the inner ring, where stress
accumulation is strong. Without viscous dissipation (bottom right
in Fig.~\ref{fig:results}) the sheet continues to expand and quickly
reaches unphysical self-intersecting configurations. In a real system,
steric repulsion and intrinsic biological processes such as apoptosis
due to hypoxia and nutrient deprivation would prevent this uncontrolled
growth. If viscoelastic relaxation is introduced, the stress generated
by active remodelling is in part dissipated, which prevents wrinkles
from growing rapidly (upper right region in Fig.~\ref{fig:results}).
The ratio between active relaxation and viscous dissipation then determines
the steady state wrinkling patterns. These patterns, however, do not
correspond to minima of elastic energy and thus exhibit far richer
morphologies compared to the equilibrium states (Fig.~S1 in \citep{SI}).

Furthermore, if the system is able to dynamically tune the active
remodelling rate, it can reach conformations that would otherwise
require ovecoming large energy barriers. For example, for a fixed
high value of $\tau_{a}^{-1}$, one needs to inject substantial energy
in order to initiate wrinkling (Fig.~\ref{fig:energy}, circles).
On the other hand, if the initial value of $\tau_{a}^{-1}$ is reduced,
the wrinkling energy barrier is significantly lowered (Fig.~\ref{fig:energy},
triangles). This is not surprising as elastic relaxation is not fast
enough to accommodate structural changes due to fast active remodelling.
If $\tau_{a}^{-1}$ is increased once the wrinkles are formed, however,
it is easy to reach different wrinkling patterns (Fig.~\ref{fig:energy}
pentagons) without the high initial energy cost. This simple example
shows that an out of equilibrium system is not only able to develop
a rich variety of morphologies but it also can avoid costly energy
barriers between different patterns by dynamically tuning its parameters,
which most biological systems are equipped to do.

\begin{figure}
\begin{centering}
\includegraphics[width=0.95\columnwidth]{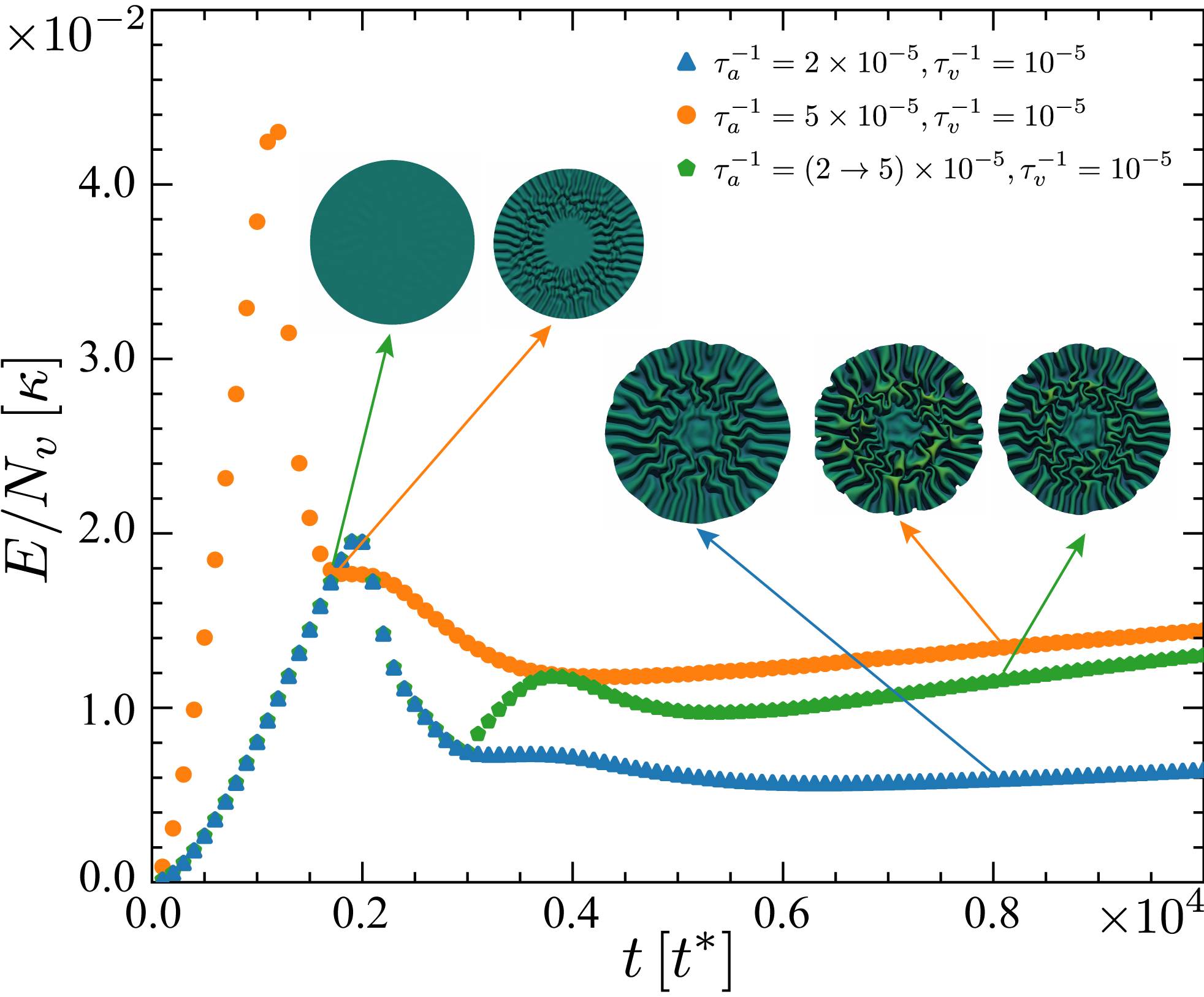}
\par\end{centering}
\caption{Energy per vertex as a function of simulation time. To reach a wrinkled
configuration with a remodelling rate $\tau_{a}^{-1}=5\times10^{-5}$
would require the sheet to overcome a large energy barrier (peak of
the orange curve). If the initial remodelling rate, however, is set
to $\tau_{a}^{-1}=2\times10^{-5}$, the system requires less energy
to reach the wrinkling instability (peak of the blue curve). Upon
switching to $\tau_{a}^{-1}=5\times10^{-5}$ at $t=1.7\times10^{4}t^{*}$,
the evolution continues along the green curve and the system reaches
a wrinkling pattern, which is very similar to the one obtained by
following the orange curve, as shown by the two snapshots on the right.
All rates are given in units of $1/t^{*}$. \label{fig:energy}}

\end{figure}

By applying an active solid model to viscoelastic thin sheets subject
to active structural remodelling, we showed that the interplay between
activity and viscous relaxation leads to a diverse morphology of out
of equilibrium wrinkling patterns. Of particular interest in this
study is the regime where active processes are faster than elastic
and viscoelastic relaxation. In this case, the system has no time
to fully relax local stresses produced by active remodelling allowing
local perturbations to grow. As a consequence, the shape patterns
depend on the initial conditions and local fluctuations. This is in
stark contrast to the mechanics of growth, in particular in plants,
that has been extensively studied with great success \citep{goriely2017mathematics}.
Most theoretical approaches are based on continuum mechanics augmented
to encode the effects of growth into F{\" o}ppl-von K{\' a}rm{\' a}n equations \citep{amar2005growth,goriely2005differential,goriely2007definition}.
The salient point in such treatments is that elastic relaxation occurs
at the time scales that are short compared to growth and thus describe
the regime where the system is always in quasi-static mechanical equilibrium
\citep{rodriguez1994stress,amar2005growth}. We argue that the out
of equilibrium regime studied here is of particular interest in developing
physical understanding of morphogenesis.

We note that a similar observation has been recently made in a study
of the dynamics of growth and form in prebiotic vesicles \citep{ruiz2019dynamics}
where the observed diversity of shapes was associated with the imbalance
of surface and volume growth and the rate of relaxation. This suggests
that keeping a growing system out of equilibrium significantly increases
the range of available morphologies. The development of higher organisms
is too complex to be captured by a simple mechanical model of actively
remodelling sheets. Our observations, however, point to a mechanism
by which a system that is kept out of equilibrium could be steered
towards a desired shape by a careful regulation of remodelling, relaxation
and mechanical parameters. This would be much easier to encode in
the space available in the genome.

RS would like to thank C.~J.~Weijer for his valuable insights into
developmental biology. FD, NSW and DMF were funded by the UK BBSRC
(Award BB/P001335/1). RS acknowledges support by the UK BBSRC (Award
BB/N009789/1). 

\bibliographystyle{apsrev4-1}

  \pagebreak
  \widetext
  \begin{center}
	\textbf{\large Supplemental Materials: Wrinkle patterns in active viscoelastic thin sheets}
  \end{center}

\setcounter{equation}{0}
\setcounter{figure}{0}
\setcounter{table}{0}
\makeatletter
\renewcommand{\theequation}{S\arabic{equation}}
\renewcommand{\thefigure}{S\arabic{figure}}
\renewcommand{\bibnumfmt}[1]{[S#1]}
\renewcommand{\citenumfont}[1]{S#1}

  
  
  \section{Energy}
  
  Here we show the elastic energy for different value of $\tau_a^{-1}$ and $\tau_v^{-1}$. Fig.\ \ref{fig:energy} shows that the elastic energy is far from the global minimum with prominent regions of highly concentrated bending energy. If viscoelastic relaxation is introduced, the stress generated by active remodelling is in part dissipated, which prevents wrinkles from growing rapidly (upper right region in Fig.\ 3 main text). The ratio between active relaxation and viscous dissipation then determines the steady state wrinkling patterns. These patterns, however, do not correspond to minima of elastic energy and thus exhibit far richer morphologies compared to the equilibrium states. It is also easy to transition between different wrinkling patterns by tuning system parameters, which most biological systems are equipped to do. Note that while the precise morphology of wrinkling patterns depends on the geometry of the system, the mechanism that leads to such out of equilibrium structures does not.
  
  \begin{figure}[ht!]
  \includegraphics[width=0.5\textwidth]{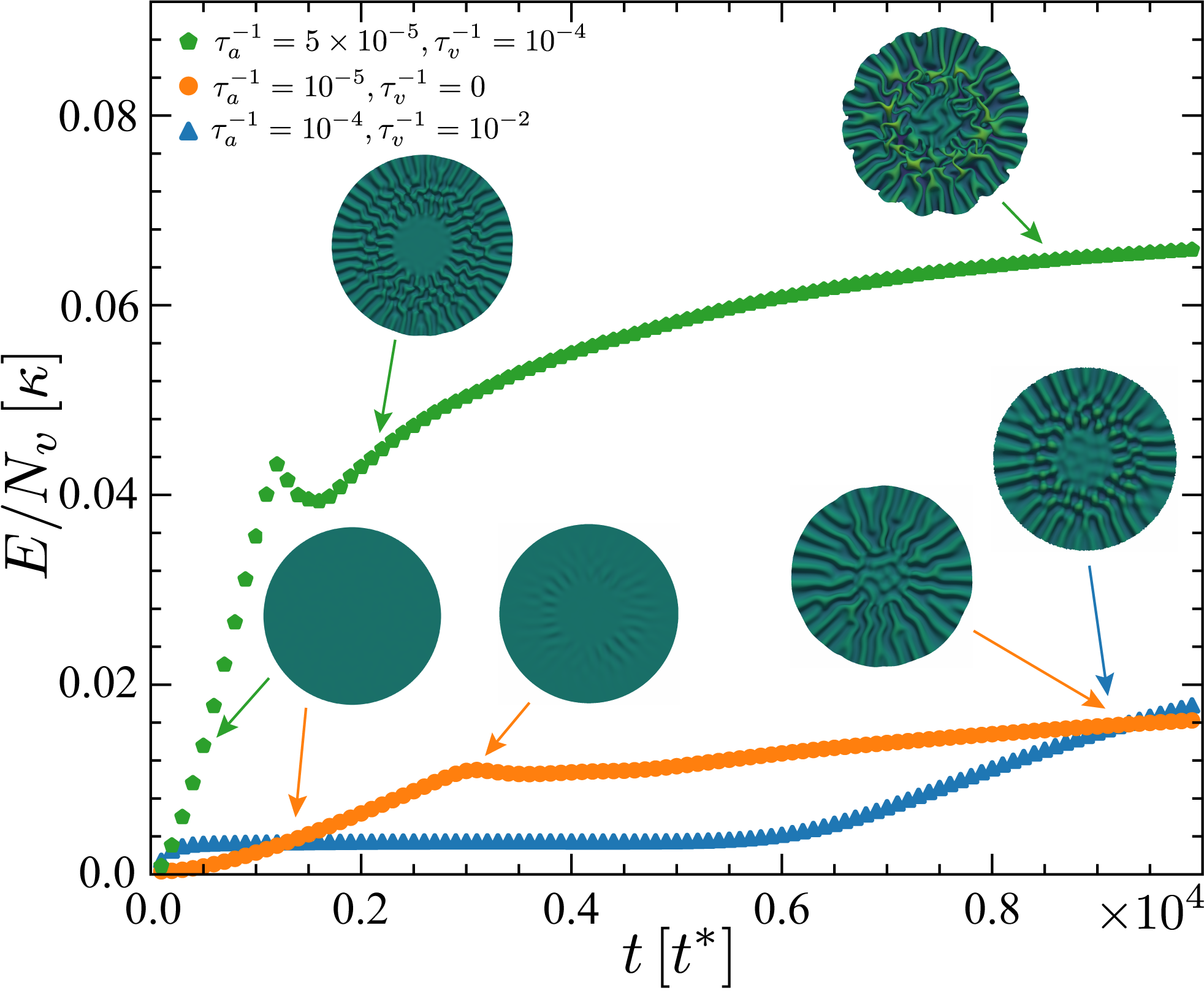}
  \caption{Total elastic energy $E$ divided by the number of vertices $N_{v}$ as a function of the simulation time. Note that small peaks in the green and orange curves correspond to the onset or wrinkling. The sheet represented by the blue curve wrinkles at around $t=60t^{*}$, however, there is no distinct peak due to very strong viscous relaxation. Once the wrinkles form, the energy gradually increases due to active remodelling. Note that within our model, in most cases, the system would not reach a steady state and different mechanism would have to be introduced to stabilise the system. $\tau_{a}^{-1}$ and $\tau_{v}^{-1}$ are measured in units of $t^{*-1}$. }
  \label{fig:energy}
  \end{figure}

  \section{Other geometries and remodelling tensors}
\begin{figure*}[ht!]
  \includegraphics[width=0.5\textwidth]{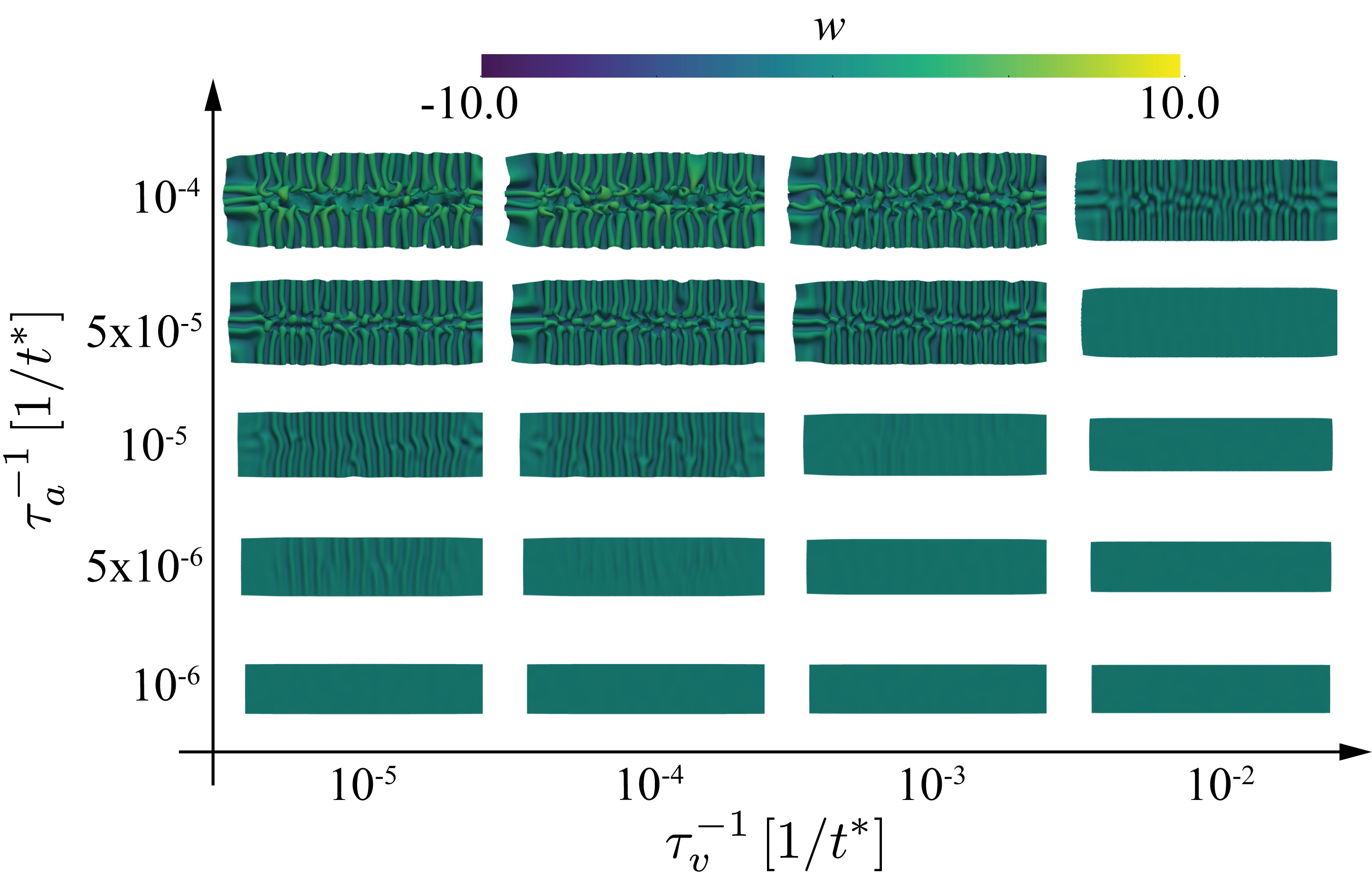}
  \caption{A snapshot of the out-of equilibrium shapes obtained by numerical integration of Eqs.\ (2) and (3) in the main text starting from a flat stripe configuration. The snapshots are taken at $t=4\times 10^{3}t^{*}$. Horizontal axis represents the rate of viscous (dissipative) relaxation with increasing values designating faster residual stress relaxation. On the vertical axis we plot the active structural remodelling rate, with larger values corresponding to faster changes of the local reference metric. The usual slow, quasi-equilibrium elastic growth would correspond to the lower left corner in this graph. Colours represent the height function, $w\left(x,y\right)$.}
  \label{fig:strip}
  \end{figure*}

  Here we show two examples of geometries and structural remodelling. The first example (Fig.\ \ref{fig:strip}) shows the case of a strip of size $L_x = 100$ and $L_y=20$ under uniform structural and viscous remodelling. As in the case of Fig.\ 3 in the main text, if one increase the rate of viscoelastic relaxation while keeping the active remodelling wrinkles are less
  pronounced or, in the case of very fast dissipative relaxation, do not form at all.
  
\begin{figure*}[ht!]
  \includegraphics[width=0.75\textwidth]{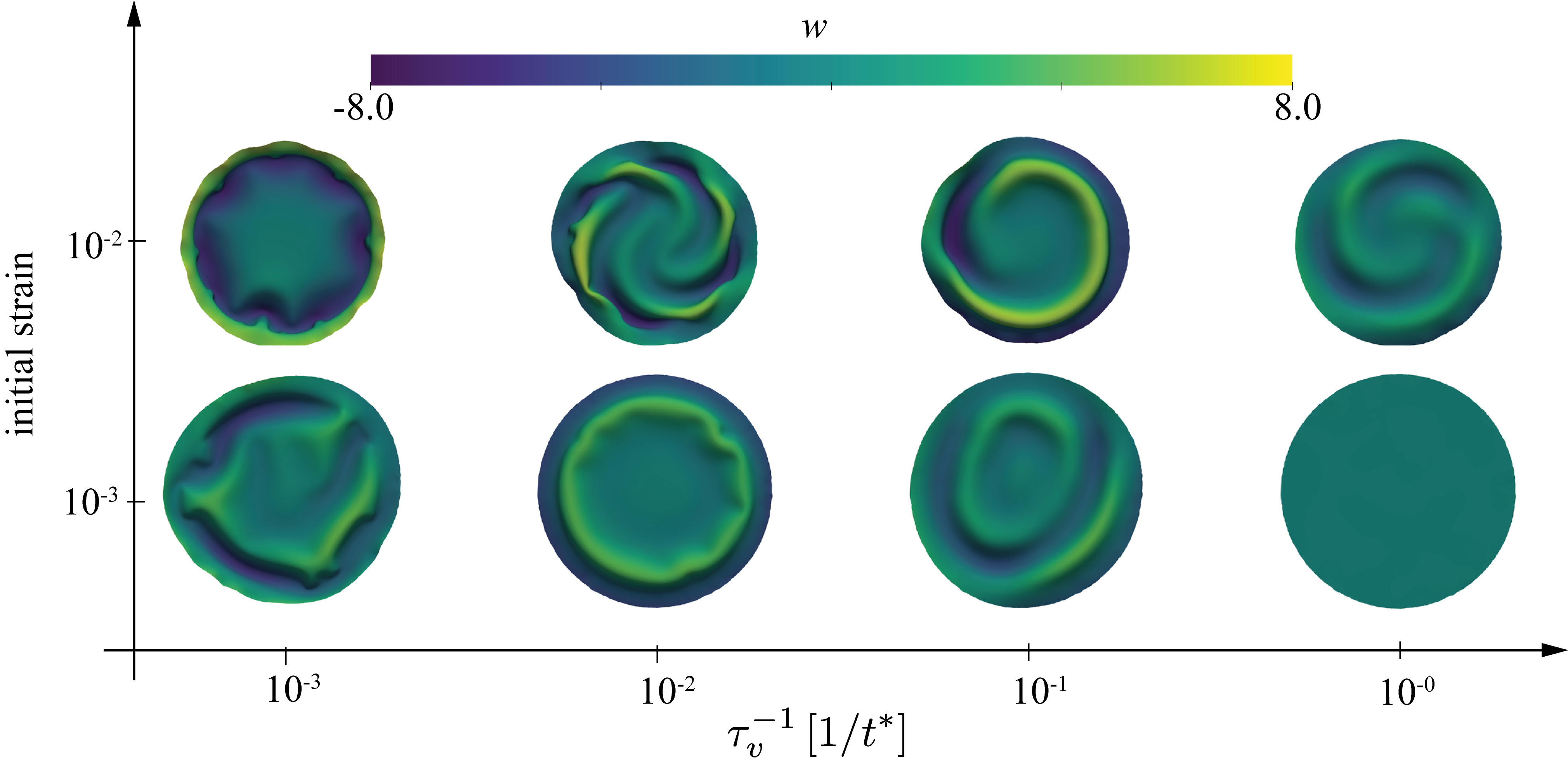}
  \caption{A snapshot of the out-of equilibrium shapes obtained by numerical integration of Eqs.\ (2) and (3) in the main text starting from a flat stripe configuration. The snapshots are taken at $t=2\times 10^{4}t^{*}$. Horizontal axis represents the rate of viscous (dissipative) relaxation with increasing values designating faster residual stress relaxation. On the vertical axis, we plot the initial residual strain in the exterior annulus, with larger values corresponding to larger changes of the local reference metric. Colours represent the height function, $w\left(x,y\right)$.}
  \label{fig:compression}
  \end{figure*}

  The second example (Fig.\ \ref{fig:compression}), mimics active compression of a flat disk of size $R=50$. The compression is introduced by imposing a rapid strain trough an instantaneous change on the reference metric in an external annulus $0.8R<r<R$. Viscous remodelling is assumed to occur only in the inner annulus, for $r<0.8R$.

  \section{Elasticity}
  
\begin{figure}[h!]
  \begin{centering}
  \includegraphics[scale=0.6]{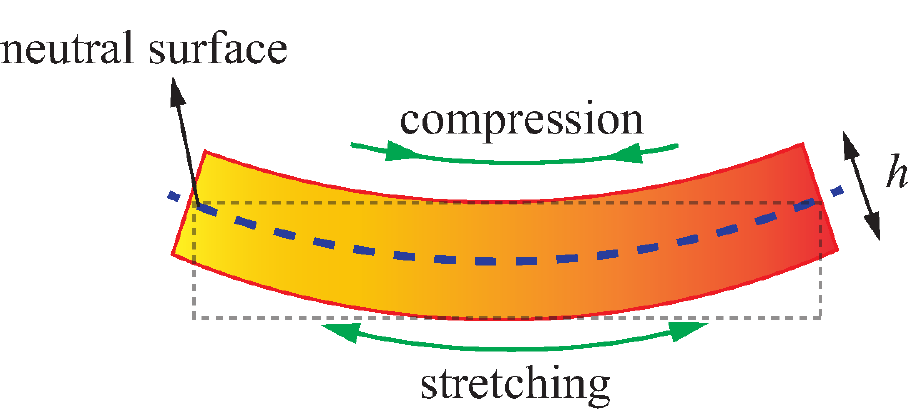}
  \par\end{centering}
  \caption{A perpendicular cut through a thin sheet of thickness $h\ll1$. As
  the sheet is bend upwards from its rest configuration (dashed black
  lines) it deforms. The inner side (towards the direction of the bend)
  gets compressed while the outer side becomes stretched. It is clear
  that as one moves away from the inner side towards the outer side
  the amount of compression decreases and eventually turns into stretching.
  Therefore, there is a surface which neither stretches nor compresses, the so-called \emph{neutral surface}
   (blue dashed line) and is clearly exactly in the middle of the sheet.\label{fig:neutral_surface}}
  \end{figure}

  Starting for the energy expression for a thin three-dimensional solid
  we derive energy expression for its two-dimensional neutral surface.
  The neutral surface is placed midway along the thin direction (see Fig.\ \ref{fig:neutral_surface}) and on it
  bending and stretching are decoupled. 
  
  \subsection{Strain tensor}
  
  A point at position $\mathbf{r}$ under the deformation
  is displaced to \citep{Green}
  \begin{equation}
  \mathbf{r}^\prime=\mathbf{r}+\mathbf{u},\label{eq:displacement_vector}
  \end{equation}
  where vector $\mathbf{u}\left(\mathbf{r}\right)$ is the displacement vector.
  The elastic energy of the body cannot depend
  on actual displacement but it depends on the derivatives of $\mathbf{u}$. In other words,
  the elastic energy depends on the changes in the metric of the body.
  We define the \emph{strain tensor} 
  \begin{equation}
      u_{ij}=\frac{1}{2}\left(g_{ij}-\bar{g}_{ij}\right).\label{eq:strain_general}
  \end{equation}
  If we recall that
  \begin{equation*}
  \bar{g}_{ij}=\partial_{i}\mathbf{r}\cdot\partial_{j}\mathbf{r}=\mathbf{e}_{i}\cdot\mathbf{e}_{j},
  \end{equation*}
  and
  \begin{equation*}
  \begin{split}
      g_{ij} & = \bar{g}_{ij}+\mathbf{e}_{i}\cdot\partial_{j}\mathbf{u}+\mathbf{e}_{j}\cdot\partial_{i}\mathbf{u}+\partial_{i}\mathbf{u}\cdot\partial_{j}\mathbf{u},
  \end{split}
  \end{equation*}
  the strain tensor becomes
  \begin{equation}
  u_{ij}=\frac{1}{2}\left(\nabla_{j}u_{i}+\nabla_{i}u_{j}+\nabla_{i}u_{s}\nabla_{j}u^{s}\right),\label{eq:strain_displacement_curvilinear}
  \end{equation}
  where $\nabla_{j}$ is the covariant derivative.
  In the Euclidean space, $\nabla_{i}=\partial_{i}$, and the last expression
  reduces to the familiar definition of the strain tensor defined in
  standard text books on elasticity
  \begin{equation}
  u_{ij}=\frac{1}{2}\left(\partial_{i}u_{j}+\partial_{j}u_{i}+\partial_{i}u_{s}\partial_{j}u^{s}\right).\label{eq:strain_displacement_flat_space}
  \end{equation}
  
  \subsection{Three-dimensional elastic energy density}
  
  The elastic energy density depends on the strain tensor, i.e., on the metric, $E_{el}=E_{el}\left(g_{ij}\right)$.
  If the strain is small we can expand $E_{el}$ in powers of $u_{ij}$
  around the reference configuration ($\bar{g}_{ij}$ and $u_{ij}=0$)
  \begin{equation*}\begin{split}
  E_{el} & \approx  E\left(\bar{g}_{ij}\right)+\left.\frac{\partial E}{\partial g_{ij}}\right|_{u_{ij}=0}u_{ij}+\frac{1}{2}\left.\frac{\partial^{2}E}{\partial g_{ij}\partial g}\right|_{u_{ij}=0}u_{ij}u_{kl}\\
  &+o\left(u^{3}\right)\\
   & =  E\left(\bar{g}_{ij}\right)+\frac{1}{2}A^{ijkl}u_{ij}u_{kl}+o\left(u^{3}\right),
  \end{split}\end{equation*}
  where we assumed the linear term in the expansion vanishes and introduced
  a contravariant elastic tensor $A^{ijkl}=\left.\frac{\partial^{2}E}{\partial g_{ij}\partial g_{kl}}\right|_{u_{ij}=0}$.
  We can omit the unimportant constant term $E\left(\bar{g}_{ij}\right)$
  to obtain the expression for elastic energy density in the small strain
  approximation
  \begin{equation}
  E_{el}=\frac{1}{2}A^{ijkl}u_{ij}u_{kl}.\label{eq:E_elasitc}
  \end{equation}
  We need to make a distinction between small strain
  and small displacement approximations, that is, one does not imply
  the other. If the strains are small and the elastic response of the
  body is directly proportional to the applied stress, the small strain
  approximation is applicable and Eq. (\ref{eq:E_elasitc}) is valid,
  i.e. constitutive laws are linear (Hookean elasticity). However, when studying thin shells,
  we can have a situation that although strains are small displacements
  are large. In this case Eq. (\ref{eq:E_elasitc}) is still valid,
  but we cannot omit nonlinear terms in $u_{ij}$ and we have to use
  Eq. (\ref{eq:strain_displacement_curvilinear}). 
  
  Total elastic energy is \cite{Koiter,Efrati},
  \begin{equation}
  E_{tot}=\frac{1}{2}\int_{V}\sqrt{\left|g\right|}A^{ijkl}u_{ij}u_{kl}.\label{eq:E_total}
  \end{equation}
  For an isotropic body there the elastic tensor has only two independent
  components and can be written as (e.g., Ref. \citep{Efrati})
  \begin{equation*}
  A^{ijkl}=\lambda g^{ij}g^{kl}+\mu\left(g^{ik}g^{jl}+g^{il}g^{jk}\right),
  \end{equation*}
  where $\lambda$ and $\mu$ are two Lame coefficients. We can introduce
  Young's modulus $E$ and Poisson's ratio $\nu$ via
  \begin{equation}\begin{split}
  E & =  \frac{\mu\left(3\lambda+2\mu\right)}{\lambda+\mu}\nonumber \\
  \nu & =  \frac{\lambda}{2\left(\lambda+\mu\right)}. \label{eq:Young_Lame}
  \end{split}
  \end{equation}

  \subsection{Two-dimensional plate Energy density}
  
  Expression of the elastic energy density of the
  neutral surface can be derived \citep{Koiter2} under the Kirchhoff-Love
  assumptions (Refs.\ \citep{Koiter2} and \citep{Efrati}):
  \begin{enumerate}
  \item Body is in the state of plane-stress, i.e., stress normal to the surfaces
  parallel to the neutral surface can be neglected.
  \item Points which lie on a normal to the neutral surface in the reference
  configuration remain on the same normal in the deformed configuration.
  \end{enumerate}
  These assumptions translate into
  \begin{equation}
  \sigma^{i3}=0\,\,\,\,\,\,i=x,y,z\label{eq:Kirchhoff_1}
  \end{equation}
  where $\sigma^{ij}$ is the contravariant stress tensor, and 
  \begin{equation}
  g_{ij}=\left(\begin{array}{cc}
  g_{\alpha\beta} & 0\\
  0 & 1
  \end{array}\right)\,\,\,\,\mathrm{or\,\,\,\,\epsilon_{\alpha3}=0.}\label{eq:Kirchhoff_2}
  \end{equation}
  From Eqs.\ (\ref{eq:Kirchhoff_1}) and (\ref{eq:Kirchhoff_2})
  we have 
  \begin{equation}
  u_{3}^{3}=u_{33}=-\frac{\lambda}{\lambda+2\mu}u_{\alpha}^{\alpha}.\label{eq:u33}
  \end{equation}
  The elastic energy density in Eq.\ (\ref{eq:E_elasitc}) can be now
  rewritten as
  \begin{equation*}\begin{split}
  E_{tot}^{2D} & =  \frac{1}{2}\mathcal{A}^{\alpha\beta\gamma\delta}u_{\alpha\beta}u_{\gamma\delta},
  \end{split}\end{equation*}
  where the two-dimensional elastic tensor is
  \begin{equation}
  \mathcal{A}^{\alpha\beta\gamma\delta}=2\mu\left(\frac{\lambda}{\lambda+2\mu}g^{\alpha\beta}g^{\gamma\delta}+g^{\alpha\gamma}g^{\beta\delta}\right).\label{eq:2d_elastic}
  \end{equation}
  
  Using the Kirchhoff-Love assumptions effectively decouples
  different sheets parallel to the neutral surface \citep{Koiter}. Therefore, we can
  obtain the expression for the total elastic energy of the neutral
  surface by integrating along the sheet thickness (chosen to be the
  $z$ direction), 
  \begin{equation*}\begin{split}
  E_{tot}^{2D} & =  \frac{1}{2}\int_{S}\int_{-\frac{h}{2}}^{\frac{h}{2}}dz\sqrt{\left|g\left(z\right)\right|}\mathcal{A}^{\alpha\beta\gamma\delta}u_{\alpha\beta}\left(z\right)u_{\gamma\delta}\left(z\right).
  \end{split}\end{equation*}
  In the small strain approximation, we can neglect all terms
  that are are cubic or higher power in $u_{\alpha\beta}$ to get 
  \begin{equation}
  E_{tot}^{2D}=\int_{S}\sqrt{\left|g\right|}\mathcal{A}^{\alpha\beta\gamma\delta}\left(\frac{h}{2}u_{\alpha\beta}u_{\gamma\delta}+\frac{h^{3}}{24}b_{\alpha\beta}b_{\gamma\delta}\right),\label{eq:E_shell}
  \end{equation}
  where $b_{ij} = \mathbf{e_i}\cdot\partial_j\mathbf{n}$ is the second fundamental form, related to the curvature tensor $c_i^j=g^{ik}b_{kj}$ \citep{doCarmo}.
  Eq.\ (\ref{eq:E_shell}) is the expression for the elastic energy of
  a thin shell expressed in terms of its neutral surface. The first
  term in the two-dimensional energy expression is stretching energy
  and it describes energy penalty of stretching or compressing of the
  neutral surface. The second term is the bending energy, which describes energy penalty of flexing the sheet. We can write $E^{2D}=E_{stretch}+E_{bend}$ where
  \begin{equation}
  E_{stretch} = \frac{Y}{2\left(1+\nu\right)}\left(\frac{\nu}{1-\nu}u_{\alpha}^{\alpha}u_{\beta}^{\beta}+u_{\beta}^{\alpha}u_{\alpha}^{\beta}\right),\label{eq:E_stretch}
  \end{equation}
  and $Y=Eh$. Similarly, using $\mathrm{Tr}\left(b_{\alpha}^{\beta}\right)=2H$
  and $\mathrm{Tr}\left(\left(b_{\mu}^{\gamma}\right)^{2}\right)=4H^{2}-2K$, 
  \begin{equation}
  \begin{split}
  E_{bend} & =  \frac{h^{3}}{24}\frac{E}{\left(1+\nu\right)}\left(\frac{\nu}{1-\nu}b_{\alpha}^{\alpha}b_{\gamma}^{\gamma}+b_{\alpha}^{\beta}b_{\beta}^{\alpha}\right) \\
   & =  2\kappa H^{2}+\kappa_{G}K,
  \end{split}
  \label{eq:E_bend}
  \end{equation}
  with the bending modulus, $\kappa=Eh^{3}/12\left(1-\nu^{2}\right)$ and Gaussian modulus $\kappa_{G}=-Eh^3/12\left(1+\nu\right)$.
  
  \section{Discrete model}
  
  \begin{figure}[h!]
  \begin{centering}
  \includegraphics[scale=0.3]{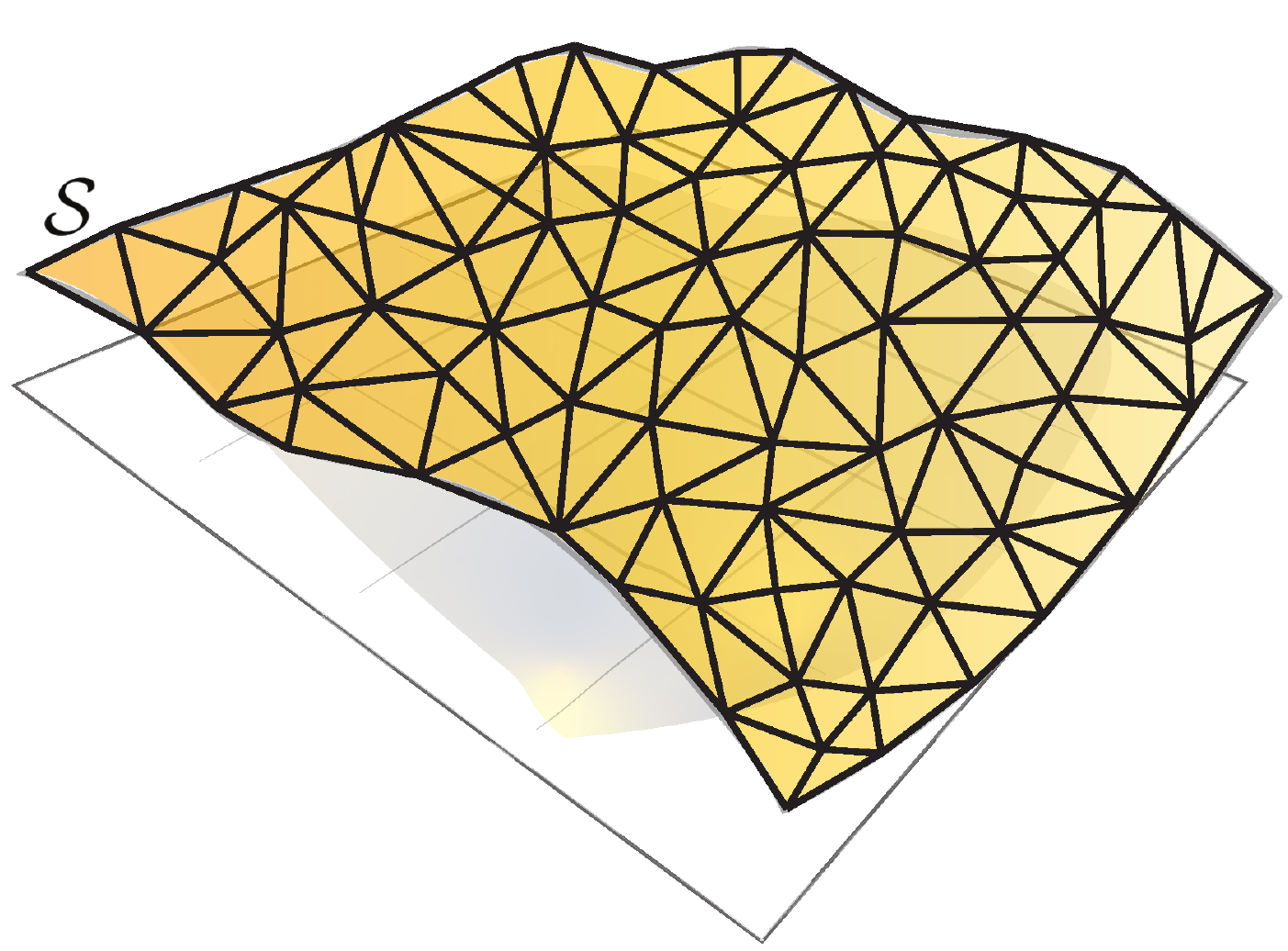}\includegraphics[scale=0.3]{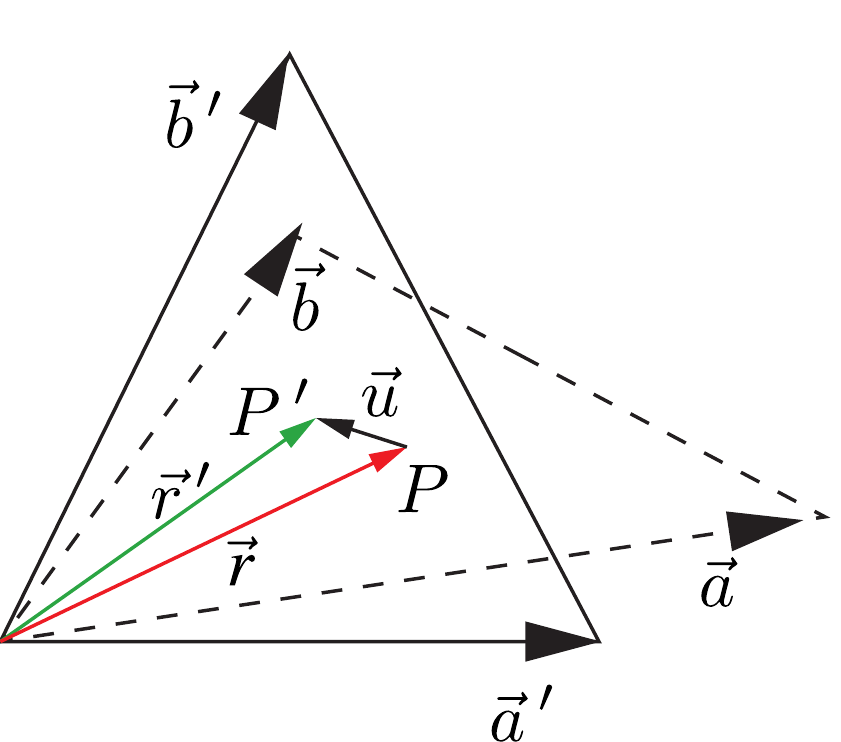}
  \par\end{centering}
  \caption{(Left) Discrete model of a thin sheet modelled as a two-dimensional
  surface. The surface is represented as a set of vertices connected
  by edges to form triangles. (Right) A point $P$ described by vector
  $\mathbf{r}$ in the undeformed (reference) triangle spanned by vectors
  $\mathbf{a}$ and $\mathbf{b}$ is moved to the point $P\,'$ with vector
  $\mathbf{r}\,'$ in the deformed triangle, spanned by vectors $\mathbf{a}'$
  and $\mathbf{b}'$.\label{fig:triangle_deformation} \label{fig:surface_model} }
  \end{figure}
  We discretise the surface using a triangular mesh \citep{Seung}, Fig.\ \ref{fig:surface_model}.
  
  \subsection{Stretching energy}
  
  We start with the stretching energy term. Closely following
  Ref.\ \citep{Parrinello}, coordinates of a given
  point $P$ inside a triangle can be
  written in terms of the two vectors $\mathbf{a}$ and $\mathbf{b}$ (Fig. \ref{fig:triangle_deformation}),
  \begin{equation}
  \mathbf{r}=\xi\mathbf{a}+\eta\mathbf{b},\label{eq:r_in_triangle}
  \end{equation}
  where $0\le\xi\le1$ and $0\le\eta\le1$ are coordinates of vector
  $\mathbf{r}$ in the basis $\left\{ \mathbf{a},\mathbf{b}\right\}$. Note that vectors $\mathbf{a}$ and $\mathbf{b}$ are themselves three
  dimensional vectors constructed as 
  \begin{equation*}\begin{split}
  \mathbf{a} & =  \mathbf{r}_{2}-\mathbf{r}_{1},\\
  \mathbf{b} & =  \mathbf{r}_{3}-\mathbf{r}_{1},
  \end{split}\end{equation*}
  where $\mathbf{r}_{1}$, $\mathbf{r}_{2}$ and $\mathbf{r}_{3}$ are positions of the three corners of the triangle. 
  For convenience, that is to be able to work with square matrices that are invertible,
  we will introduce a third vector
  \begin{equation*}
  \mathbf{c}=\mathbf{a}\times\mathbf{b}.
  \end{equation*}
  We can now construct a $3\times3$ matrix 
  \begin{equation}
  \hat{h}=\left(\begin{array}{cc}
  \mathbf{a} & \mathbf{b}\end{array}\right)\equiv\left(\begin{array}{ccc}
  a_{x} & b_{x} & c_{x}\\
  a_{y} & b_{y} & c_{y}\\
  a_{z} & b_{z} & c_{z}
  \end{array}\right),\label{eq:h_matrix}
  \end{equation}
  such that
  \begin{equation}
  \mathbf{r}=\hat{h}\mathbf{s},\label{eq:r_h_vec}
  \end{equation}
  where 
  \begin{equation*}
  \mathbf{s}=\left(\begin{array}{ccc}
  \xi & \eta & \omega\end{array}\right)^{T},
  \end{equation*}
  where $\omega\equiv0$. In term of coordinates Eq.\ (\ref{eq:r_h_vec})
  can be written as
  \begin{equation*}
  r_{i}=h_{i}^{\alpha}s_{\alpha},
  \end{equation*}
  and we have used Latin indices to count components in the embedding
  space and Greek indices to count components of vector $\mathbf{s}$.
  If we now introduce point $Q$ in the same triangle with coordinates
  $\phi$ and $\psi$, i.e., with $\mathbf{s}_{Q}=\left(\begin{array}{ccc}
  \phi & \psi & 0\end{array}\right)^{T}$ then 
  \begin{equation*}
  \mathbf{r}_{Q}=\hat{h}\mathbf{s}_{Q}.
  \end{equation*}
  Square of the distance $l^{2}$ between points $P$ and $Q$ is
  \begin{equation}\begin{split}
  l^{2} & =  \left(\mathbf{r}_{P}-\mathbf{r}_{Q}\right)\cdot\left(\mathbf{r}_{P}-\mathbf{r}_{Q}\right)\nonumber \\
   & =  \left(\mathbf{s}_{\left(P\right)}-\mathbf{s}_{\left(Q\right)}\right)\hat{\bar{g}}\left(\mathbf{s}_{\left(P\right)}-\mathbf{s}_{\left(Q\right)}\right).\label{eq:triangle_dist_square}
  \end{split}\end{equation}
  and we have used parentheses in the sub- and superscript to designate
  that $P$ and $Q$ are not indices and have freely changed the name
  of the repeated summation indices. Matrix 
  \begin{equation*}
  \hat{\bar{g}}=\hat{h}^{T}\hat{h}
  \end{equation*}
  is the (discrete) metric of our reference triangle. Explicitly, 
  \begin{equation}
  \begin{split}
  \hat{\bar{g}} & =  \left(\begin{array}{ccc}
  \mathbf{a} & \mathbf{b} & \mathbf{c}\end{array}\right)^{T}\left(\begin{array}{ccc}
  \mathbf{a} & \mathbf{b} & \mathbf{c}\end{array}\right), \\
  \end{split}
  \label{eq:metric}
  \end{equation}
  i.e., matrix $\hat{\bar{g}}$ is a $3\times3$ matrix with a $2\times2$
  sub-matrix corresponding to the metric tensor of the triangle and
  the $\mathbf{c}\cdot\mathbf{c}$ term that is added for convenience. Matrix $\hat{\bar{g}}$ 
  can be easily computed. 
  
  After the deformation edges of the triangle change and the basis vectors
  become $\mathbf{a}^\prime$ and $\mathbf{b}^\prime$, while the point $P$ has moved
  to the new position $P^\prime$ with coordinates 
  \begin{equation*}
  \mathbf{R}=\hat{H}\mathbf{s}\,,
  \end{equation*}
  where $\hat{H}=\left(\begin{array}{ccc}
  \mathbf{a}^\prime & \mathbf{b}^\prime & \mathbf{c}^\prime\end{array}\right)^{T}\left(\begin{array}{ccc}
  \mathbf{a}^\prime & \mathbf{b}^\prime & \mathbf{c}^\prime\end{array}\right)$. Note that we assume that the deformation is linear 
  and as such automatically affine. An important of this restrictions is that the point $P^\prime$
  has the same coordinates $\left(\xi,\eta,0\right)$ in the deformed
  triangle $\left\{ \mathbf{a}^\prime,\mathbf{b}^\prime\right\} $ as the point $P$ had
  in the original undeformed state. Point $P^\prime$ is point $P$ after
  deformation. If we recall Eq. (\ref{eq:displacement_vector}) and
  the definition of the displacement vector
  \begin{equation*}\begin{split}
  \mathbf{u} & =  \mathbf{R}-\mathbf{r}\\
   & =  \hat{H}\hat{h}^{-1}\mathbf{r}-\mathbf{r},
  \end{split}
\end{equation*}
  where in the second line we have inverted Eq. (\ref{eq:r_h_vec})
  to get $\mathbf{r}=\hat{h}^{-1}\mathbf{s}$. Finally,
  \begin{equation}
  \mathbf{u}=\left(\hat{H}\hat{h}^{-1}-\hat{I}\right)\mathbf{r},\label{eq:displacement_discrete}
  \end{equation}
  where $\hat{I}$ is the identity matrix. In terms of coordinates we
  have
  \begin{equation*}
  u^{i}=\left[\left(\hat{H}\hat{h}^{-1}\right)_{j}^{i}-\delta_{j}^{i}\right]x^{j}.
  \end{equation*}
  In order to derive the expression for the non-linear strain tensor
  in term of matrices $\hat{H}$ and $\hat{h}$ we use Eq. (\ref{eq:strain_displacement_flat_space})
  repeated for convenience (in a form of a mixed tensor),
  \begin{equation}
  u_{i}^{j}=\frac{1}{2}\left(\frac{\partial u_{i}}{\partial x^{j}}+\frac{\partial u^{j}}{\partial x_{i}}+\frac{\partial u^{k}}{\partial x_{i}}\frac{\partial u_{k}}{\partial x^{j}}\right).\label{eq:u_def_in_discrete}
  \end{equation}
  Now we compute 
  \begin{equation*}\begin{split}
  \frac{\partial u_{i}}{\partial x^{j}} & =  \left[\left(\hat{H}\hat{h}^{-1}\right)_{i}^{k}-\delta_{i}^{k}\right]\frac{\partial x_{k}}{\partial x^{j}}\\
   & =  \left[\left(\hat{H}\hat{h}^{-1}\right)_{i}^{j}-\delta_{i}^{j}\right].
  \end{split}\end{equation*}
  If we plug the last expression into Eq. (\ref{eq:u_def_in_discrete})
  we obtain
  \begin{equation}
  \begin{split}
  \hat{u} & =  \frac{1}{2}\left[\left(\hat{H}\hat{h}^{-1}\right)^{T}\hat{H}\hat{h}^{-1}-\hat{I}\right] \\
   & =  \frac{1}{2}\left(\hat{h}^{-T}\hat{g}\hat{h}^{-1}-\hat{I}\right),\label{eq:discrete_strain_tensor_matrix_form}
  \end{split}
  \end{equation}
  where we have used $\left(AB\right)^{T}=B^{T}A^{T}$ and $A^{-T}\equiv\left(A^{-1}\right)^{T}$.
  We can now use Eq. (\ref{eq:E_stretch}) to write
  
  \begin{equation}
  E_{stretch}  =\frac{A_{T}Eh}{8(1+\nu)}\left[\frac{\nu}{1-\nu}\left(\mathrm{tr}\hat{F}\right)^{2}+\mathrm{tr}\left(\hat{F}^{2}\right)\right],
  \label{eq:discrete_stertching_energy}
  \end{equation}
  where $A_{T}$ is 
  the triangle area and the tensor $\hat{F}$ is given by,
  \begin{equation}
  \hat{F}=\hat{\bar{g}}^{-1}\hat{g}-\hat{I}.
  \label{eq:F_tensor}
  \end{equation}
  
  \subsubsection{Vertex stretching force}
  \begin{center}
  \begin{figure}
  \centering{}\includegraphics[scale=0.5]{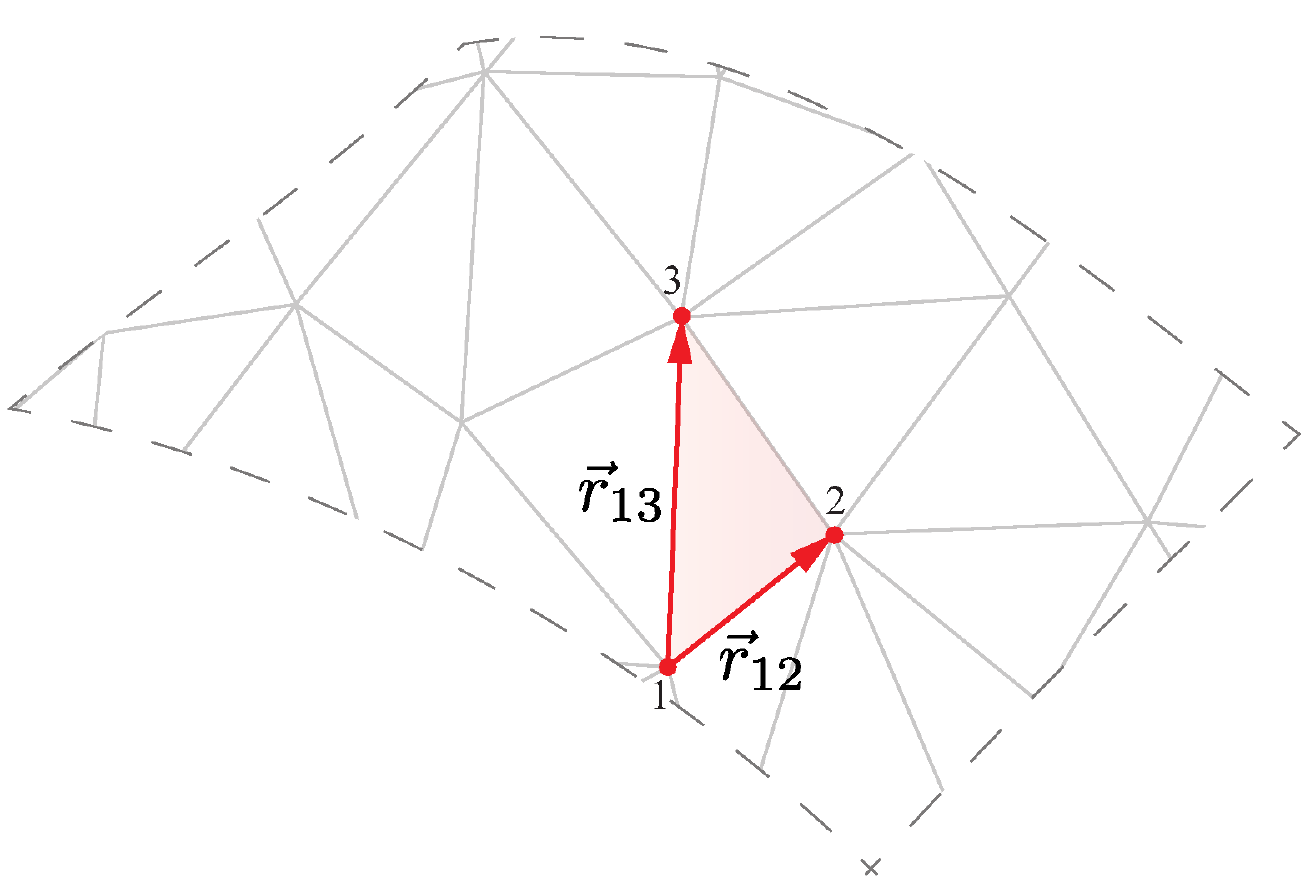}\caption{Triangular mesh showing the vector definitions for a single triangle.}
  \label{Fig:TriMesh}
  \end{figure}
  \par\end{center}
  From Eqs.\ (\ref{eq:metric}) and (\ref{eq:discrete_stertching_energy}) is easy to see that it is more convenient to
  express stretching energy in terms of the distance between vertices of one
  triangle, $\mathbf{r}_{ij}$ ($\mathbf{r}_{12},\mathbf{r}_{13}$) (Fig.\ \ref{Fig:TriMesh}). Thus, using
  the chain rule we can write the force over the triangle vertices as,
  
  \begin{equation}
  \mathbf{f}_{i}=-\left(\partial_{\mathbf{r}_{ij}}E_{s}\right)\left(\frac{\partial\mathbf{r}_{ij}}{\partial\mathbf{r}_{i}}\right).
  \end{equation}
  After lengthy but straightforward algebra, the expression for the force on vertex $i$ due to stretching reads,
  \begin{equation}
  \begin{split}\mathbf{f}_{i}= & -\left[\frac{A_{T}Yh}{4\left(1-\nu^{2}\right)}\left[\left(F_{11}+\nu F_{22}\right)\begin{pmatrix}2\alpha_{11}\mathbf{r}_{12}+\alpha_{12}\mathbf{r}_{13} & \alpha_{12}\mathbf{r}_{12}\end{pmatrix}\right.\right.\\
  + & \left(F_{22}+\nu F_{11}\right)\begin{pmatrix}\alpha_{12}\mathbf{r}_{13} & 2\alpha_{22}\mathbf{r}_{13}+\alpha_{12}\mathbf{r}_{12}\end{pmatrix}\\
  + & \left.(1-\nu)\left(F_{21}\begin{pmatrix}\alpha_{11}\mathbf{r}_{13} & 2\alpha_{12}\mathbf{r}_{13}+\alpha_{11}\mathbf{r}_{12}\end{pmatrix}+F_{12}\begin{pmatrix}2\alpha_{12}\mathbf{r}_{12}+\alpha_{22}\mathbf{r}_{13} & \alpha_{22}\mathbf{r}_{12}\end{pmatrix}\right)\right]\\
  + & \left.\frac{e_{s}}{4A_{T}}\begin{pmatrix}g_{22}\mathbf{r}_{12}-g_{12}\mathbf{r}_{13} & g_{11}\mathbf{r}_{13}-g_{12}\mathbf{r}_{12}\end{pmatrix}\right]\begin{pmatrix}-1 & 1 & 0\\
  -1 & 0 & 1
  \end{pmatrix},
  \end{split}
  \label{eq:EnergyGradient-matrix}
  \end{equation}
  where $\alpha_{ij} = \left(\bar{g}^{-1}\right)_{ij}$. This expression can be directly implemented in a simulation.

  \subsection{Bending energy}
  \begin{figure}
  \begin{centering}
  \includegraphics[scale=0.6]{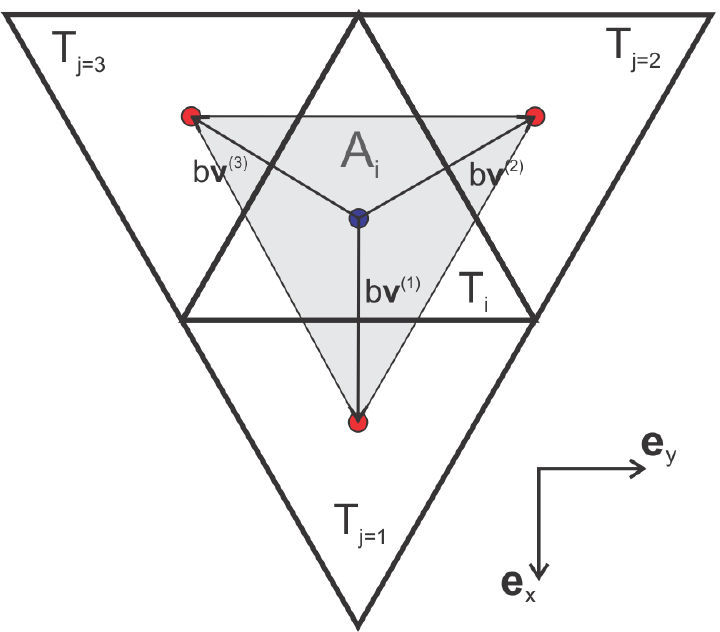}
  \par\end{centering}
  \caption{A triangle $T_{i}$ and three of its nearest neighbours $T_{j}$. In
  the continuum limit $b\to0$. Shaded are is the vertex area element.\label{fig:discrete_to_continuum}}
  \end{figure}
  In order to derive expressions of the discrete version of the bending energy we start from Eq.\ (\ref{eq:E_bend}) and use the fact that
  $\kappa_{G}=-\kappa\left(1-\nu\right)$ to obtain
  \begin{equation}
  \begin{split}
  E_{bend} & =  2\kappa H^{2}+\kappa_{G}K \\
   & =  \kappa\left(2H^{2}-K\right)+\nu\kappa K.
  \end{split}
  \label{eq:E_bend_discrete_1}
  \end{equation}
  We can now use, $\mathrm{Tr}\left[\left(b_{\mu}^{\gamma}\right)^{2}\right]=4H^{2}-2K$
  to write 
  \begin{equation*}
  \begin{split}
  E_{bend} & =  \frac{1}{2}\kappa\mathrm{Tr}\left[\left(b_{\alpha}^{\beta}\right)^{2}\right]+\nu\kappa K\\
   & =  \frac{1}{2}\kappa g^{\beta\gamma}g^{\alpha\delta}b_{\alpha\gamma}b_{\beta\delta}+\nu\kappa K.
  \end{split}
  \end{equation*}
  If we use the definition of the second fundamental form we obtain 
  \begin{equation}
  \begin{split}
  E_{bend} & =  \frac{1}{2}\kappa g^{\beta\gamma}g^{\alpha\delta}\mathbf{e}_{\alpha}\cdot\partial_{\gamma}\mathbf{n}\mathbf{e}_{\beta}\cdot\partial_{\delta}\mathbf{n}+\nu\kappa K\nonumber \\
   & =  \frac{1}{2}\kappa g^{\beta\gamma}\delta_{\beta}^{\delta}\partial_{\gamma}\mathbf{n}\cdot\partial_{\delta}\mathbf{n}+\nu\kappa K\nonumber \\
   & =  \frac{1}{2}\kappa\partial_{\gamma}\mathbf{n}\cdot\partial^{\gamma}\mathbf{n}+\nu\kappa K.
  \label{eq:E_bend_n_dot_n}
  \end{split}
  \end{equation}
  
  Following Ref.\ \citep{Seung} the $\frac{1}{2}\kappa\partial_{\gamma}\mathbf{n}\cdot\partial^{\gamma}\mathbf{n}$
  term is a continuum version of the expression
  \begin{equation}
  \begin{split}
  E_{SN}&=\frac{1}{2}\tilde{\kappa}\sum_{T_{j}.n.n.T_{i}}\left|\mathbf{n}_{T_{i}}-\mathbf{n}_{T_{j}}\right|^{2}\\
        &\underbrace{=}_{\left\Vert \mathbf{n}\right\Vert =1}\tilde{\kappa}\sum_{T_{j}.n.n.T_{i}}\left(1-\mathbf{n}_{T_{i}}\cdot\mathbf{n}_{T_{j}}\right),\label{eq:E_sn}
  \end{split}
  \end{equation}
  where $T_{i}$ is the triangle $i$, $T_{j}$ are three of its neighbours
  and $\tilde{\kappa}$ is the discrete value of the bending rigidity
  and subscript ``SN'' stands for ``Seung-Nelson''. $\tilde{\kappa}$
  is related proportional to $\kappa$ and we'll discuss the constant
  of proportionality below. The sum in Eq. (\ref{eq:E_sn}) can be written
  as
  \begin{equation*}\begin{split}
  E_{SN} & =  \frac{1}{2}\tilde{\kappa}\sum_{j}\left|\mathbf{n}\left(\mathbf{r}_{i}\right)-\mathbf{n}\left(\mathbf{r}_{i}+b\mathbf{v}^{\left(j\right)}\right)\right|^{2},
  \end{split}\end{equation*}
  where $b$ is the distance between centers of two neighbouring triangles
  and vectors $\mathbf{v}$ are (see Fig.\ \ref{fig:discrete_to_continuum})
  \begin{equation*}
  \mathbf{v}^{\left(j\right)}=\cos\left(\frac{2\pi}{3}j\right)\mathbf{e}_{x}+\sin\left(\frac{2\pi}{3}j\right)\mathbf{e}_{y}.
  \end{equation*}
  For $b\to0$ we can use Taylor series to expand $\mathbf{n}\left(\mathbf{r}_{i}+b\mathbf{v}^{\left(j\right)}\right)$
  to the linear order in $b$,
  \begin{equation}
  \mathbf{n}\left(\mathbf{r}_{i}+b\mathbf{v}^{\left(j\right)}\right)=\mathbf{n}\left(\mathbf{r}_{i}\right)+b\left.\partial^{\phi}\mathbf{n}\right|_{\mathbf{r}_{i}}v_{\phi}^{\left(j\right)}+o\left(b^{2}\right).\label{eq:n_expand}
  \end{equation}
  Thus,
  \begin{equation*}\begin{split}
  E_{SN} & =  \frac{b^{2}}{2}\tilde{\kappa}\sum_{j}\left.\partial^{\phi}\mathbf{n}\right|_{\mathbf{r}_{i}}v_{\phi}^{\left(j\right)}\left.\partial_{\psi}\mathbf{n}\right|_{\mathbf{r}_{i}}v_{\left(j\right)}^{\psi}\\
   & =  \frac{b^{2}}{2}\tilde{\kappa}\left.\partial^{\phi}\mathbf{n}\right|_{\mathbf{r}_{i}}\left.\partial_{\psi}\mathbf{n}\right|_{\mathbf{r}_{i}}\sum_{j}v_{\phi}^{\left(j\right)}v_{\left(j\right)}^{\psi}.
  \end{split}\end{equation*}
  We can now calculate the $j$-sum explicitly for each component $\phi=x,y$
  and $\psi=x,y$,
  \begin{equation*}\begin{split}
  \sum_{j=1}^{3}v_{x}^{\left(j\right)}v_{\left(j\right)}^{x} & =  \sum_{j=1}^{3}\cos^{2}\left(\frac{2\pi}{3}j\right)=\frac{3}{2}\\
  \sum_{j=1}^{3}v_{y}^{\left(j\right)}v_{\left(j\right)}^{y} & =  \sum_{j=1}^{3}\sin^{2}\left(\frac{2\pi}{3}j\right)=\frac{3}{2}\\
  \sum_{j=1}^{3}v_{x}^{\left(j\right)}v_{\left(j\right)}^{y} & =  \sum_{j=1}^{3}\sin\left(\frac{2\pi}{3}j\right)\cos\left(\frac{2\pi}{3}j\right)=0,
  \end{split}\end{equation*}
  and
  \begin{equation*}
  E_{SN}=\frac{3}{2}\frac{b^{2}}{2}\tilde{\kappa}\partial^{\phi}\mathbf{n}\partial_{\phi}\mathbf{n},
  \end{equation*}
  and we assume that $\partial_{\phi}\mathbf{n}$ is calculated at point
  $\mathbf{r}_{i}$. We write 
  $E_{SN}=-\frac{3}{2}\frac{b^{2}}{2}\tilde{\kappa}\partial^{\phi}\mathbf{n}b_{\phi}^{\mu}\mathbf{e}_{\mu}=\frac{3}{2}\frac{b^{2}}{2}\tilde{\kappa}\mathrm{Tr}\left(b_{\phi}^{\mu}\right)^{2}=\frac{3b^{2}}{4}\tilde{\kappa}\left(4H^{2}-2K\right)=\frac{3b^{2}}{2}\tilde{\kappa}\left(2H^{2}-K\right)$. Thus,
  the total discrete energy is
  \begin{equation*}
  E_{SN}^{tot}=\frac{3}{2}\tilde{\kappa}\sum_{T_{i}}b^{2}\left(2H_{T_{i}}^{2}-K_{T_{i}}\right),
  \end{equation*}
  where $H_{T_{i}}$ and $K_{T_{i}}$ are mean and Gaussian curvature
  of the triangle $T_{i}$ and the sum goes over all triangles. From
  Fig.\ \ref{fig:discrete_to_continuum} we see that the area element $A_{i}=\frac{b^{2}\sqrt{3}}{2}$,
  which leads to
  \begin{equation*}
  E_{SN}^{tot}  =  \frac{1}{\sqrt{3}}\tilde{\kappa}\sum_{T_{i}}A_{i}\left(2H_{T_{i}}^{2}-K_{T_{i}}\right),
  \end{equation*}
  which in the limit $A_{i}\to0$ becomes an integral
  \begin{equation*}
  E_{SN}^{tot}=\frac{1}{\sqrt{3}}\tilde{\kappa}\int_{A}\left(2H^{2}-K\right).
  \end{equation*}
  Comparing Eq.\ (\ref{eq:E_bend_discrete_1}) with the last expression
  we obtain $\kappa=\frac{1}{\sqrt{3}}\tilde{\kappa}$.
  Note that this is a different prefactor than obtained in Refs.\ \citep{Seung}
  and \citep{Schmidt}. The reason is that in Eq.\ (\ref{eq:n_expand})
  we have truncated the expansion to early. The exact constant of proportionality
  is of the same order of magnitude and is given as $\kappa=\frac{\sqrt{3}}{2}\tilde{\kappa}$
  Therefore, we have showed that Eq.\ (\ref{eq:E_sn}) is a good discrete approximation for the continuum elastic energy.
  The advantage of the last expression is that it can be easily computed
  in a simulation.
  
  \subsubsection{Vertex bending force}
  \begin{center}
  \begin{figure}[h!]
  \begin{centering}
  \includegraphics[scale=0.5]{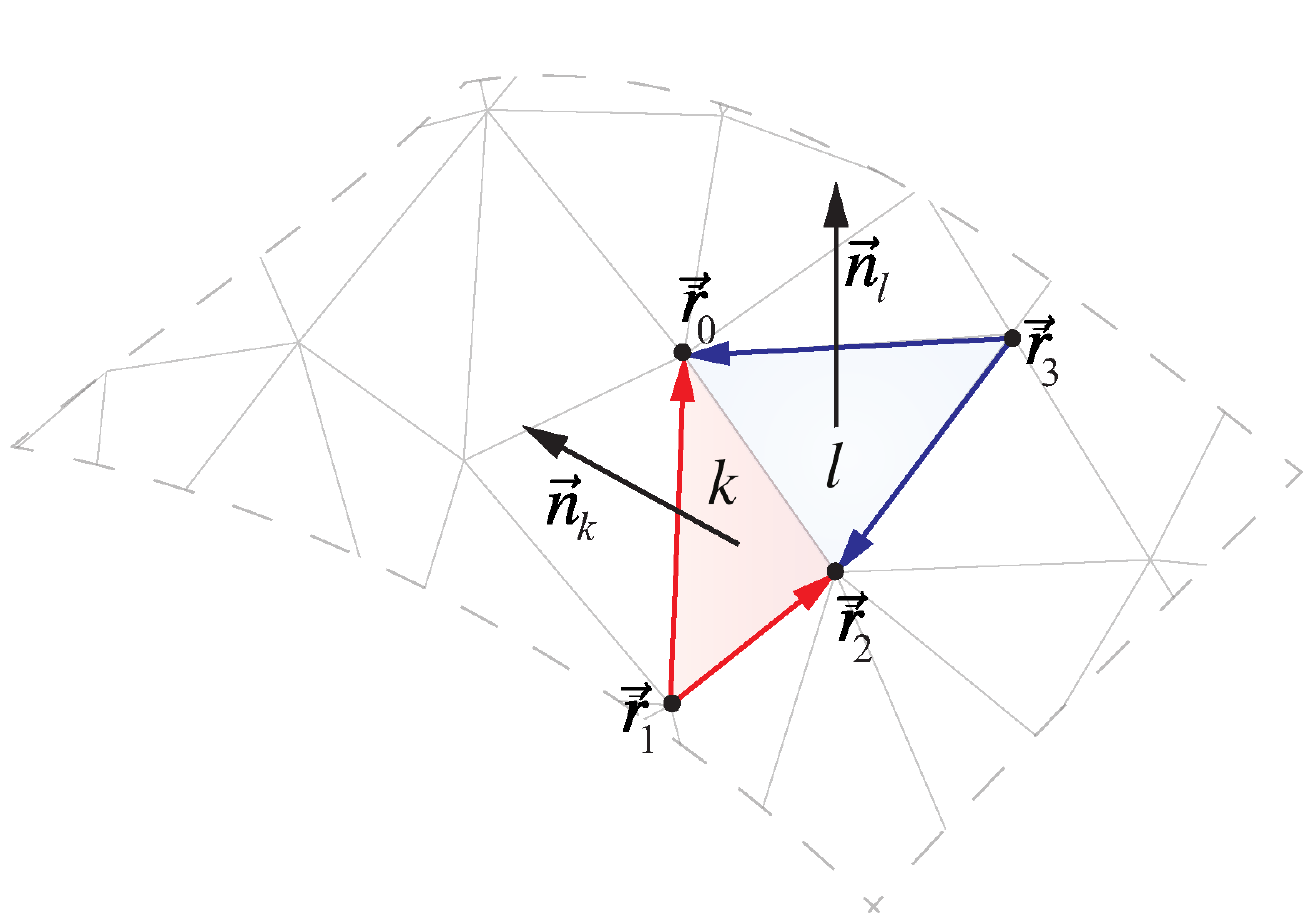}
  \par\end{centering}
  \caption{Edge $e$ shared by the faces $k$ and $l$ shown along with their
  associated normals.}
  \label{Fig:Bending_mesh}
  \end{figure}
  \par\end{center}
  
  The bending energy of a triangular mesh can be expressed as Eq.\ (\ref{eq:E_sn}),
  
  \begin{equation}
  E_{B}=\tilde{\kappa}\sum_{e\in \text{Edges}}\left(1-{\mathbf{n}}_{e,1}\cdot{\mathbf{n}}_{e,2}\right),\label{eq:bending_energy}
  \end{equation}
  where $e$ is an edge of the mesh and $\mathbf{n}_{e,1(2)}$ are the unit-length normals of two neighbouring triangles. 
  For the shake of simplicity we consider a sole edge and two triangles
  $k$ and $l$ as is shown in Fig.\ \ref{Fig:Bending_mesh}. Note that $\mathbf{n}_{k(l)}=\mathbf{A}_{k(l)}/A_{k(l)}$, 
  where $\mathbf{A}_{k(l)}$ is a vector normal to the triangle having length equal to the triangle's area.
  
  Now the bending force over a vertex $p$ is the negative gradient
  $E_B$ calculated at $\mathbf{r}_{p}$,
  \begin{equation}
  \begin{split}\mathbf{f}_{p,e} & =-\tilde{\kappa}\nabla_{\mathbf{r}_{p}}\sum_{e\epsilon Edges}\left(\frac{\mathbf{A}_{k}\cdot\mathbf{A}_{l}}{\vert\mathbf{A}_{k}\vert\,\vert\mathbf{A}_{l}\vert}\right)\\
   & =-\tilde{\kappa}\nabla_{\mathbf{r}_{p}}\sum_{e\epsilon Edges}\mathbf{f}_{p,e},
  \end{split}
  \label{eq:bending_force}
  \end{equation}
  is easy to see that since we are dealing with a sum over all the edges
  then it suffice to calculate the gradient $\nabla_{\mathbf{r}_{p}}$
  for one generic edge, 
  \begin{equation}
  \mathbf{f}_{p,e}=-\tilde{\kappa}\left(\partial_{\mathbf{r}_{ij}}E_{B,e}\right)\left(\frac{\partial\mathbf{r}_{ij}}{\partial\mathbf{r}_{p}}\right).\label{eq:bending_force_jacobian}
  \end{equation}
  After straightforward but lengthy algebra, the force matrix $\mathbf{f}_{p,e}$, Eq.\ (\ref{eq:bending_force_jacobian})
  is,
  \begin{equation}
  \mathbf{f}_{p,e}=-\tilde{\kappa}\left(\partial_{\mathbf{r}_{ij}}E_{B,e}\right)\begin{pmatrix}-1 & 1 & 0 & 0\\
  -1 & 0 & 1 & 0\\
  -1 & 0 & 0 & 1
  \end{pmatrix},
  \end{equation}
  with
  \begin{equation}
  \begin{split}\left(\partial_{\mathbf{r}_{ij}}E_{B,e}\right)= & \partial_{\mathbf{r}_{ij}}\left(\frac{\mathbf{A}_{k}\cdot\mathbf{A}_{l}}{\vert\mathbf{A}_{k}\vert\,\vert\mathbf{A}_{l}\vert}\right)\\
  = & \frac{1}{\vert\mathbf{A}_{k}\vert\,\vert\mathbf{A}_{l}\vert}\left[\partial_{\mathbf{r}_{ij}}\left(\mathbf{A}_{k}\cdot\mathbf{A}_{l}\right)-\left(\mathbf{A}_{k}\cdot\mathbf{A}_{l}\right)\left(\frac{1}{\vert\mathbf{A}_{k}\vert}\partial_{\mathbf{r}_{ij}}\vert\mathbf{A}_{k}\vert+\frac{1}{\vert\mathbf{A}_{l}\vert}\partial_{\mathbf{r}_{ij}}\vert\mathbf{A}_{l}\vert\right)\right]\\
  = & \frac{1}{4\vert\mathbf{A}_{k}\vert\,\vert\mathbf{A}_{l}\vert}\left[\begin{array}{c}
  \\
  \\
  \end{array}\right.\\
   & \left(\left(\mathbf{r}_{02}\cdot\mathbf{r}_{03}\right)\mathbf{r}_{02}-\left(\mathbf{r}_{02}\cdot\mathbf{r}_{02}\right)\mathbf{r}_{03}\left(\mathbf{r}_{02}\cdot\mathbf{r}_{03}\right)\mathbf{r}_{01}\right.\\
   & +\left(\mathbf{r}_{01}\cdot\mathbf{r}_{02}\right)\mathbf{r}_{03}-2\left(\mathbf{r}_{01}\cdot\mathbf{r}_{03}\right)\mathbf{r}_{02}\left(\mathbf{r}_{01}\cdot\mathbf{r}_{02}\right)\mathbf{r}_{02}\\
   & -\left.\left(\mathbf{r}_{02}\cdot\mathbf{r}_{02}\right)\mathbf{r}_{01}\right)\\
  + & \left(\mathbf{A}_{k}\cdot\mathbf{A}_{l}\right)\left(\frac{1}{\vert\mathbf{A}_{k}\vert^{2}}\begin{pmatrix}\left(\mathbf{r}_{02}\cdot\mathbf{r}_{02}\right)\mathbf{r}_{01}-\left(\mathbf{r}_{01}\cdot\mathbf{r}_{02}\right)\mathbf{r}_{02} & \left(\mathbf{r}_{01}\cdot\mathbf{r}_{01}\right)\mathbf{r}_{02}-\left(\mathbf{r}_{01}\cdot\mathbf{r}_{02}\right)\mathbf{r}_{01} & 0\end{pmatrix}\right.\\
  + & \left.\left.\frac{1}{\vert\mathbf{A}_{l}\vert^{2}}\begin{pmatrix}0 & \left(\mathbf{r}_{03}\cdot\mathbf{r}_{03}\right)\mathbf{r}_{02}-\left(\mathbf{r}_{02}\cdot\mathbf{r}_{03}\right)\mathbf{r}_{03} & \left(\mathbf{r}_{02}\cdot\mathbf{r}_{02}\right)\mathbf{r}_{03}-\left(\mathbf{r}_{02}\cdot\mathbf{r}_{03}\right)\mathbf{r}_{02}\end{pmatrix}\right)\right].
  \end{split}
  \end{equation}
  
  \section{Active Remodelling}
  
\begin{figure}[h!]
      \begin{centering}
      \includegraphics[scale=0.35]{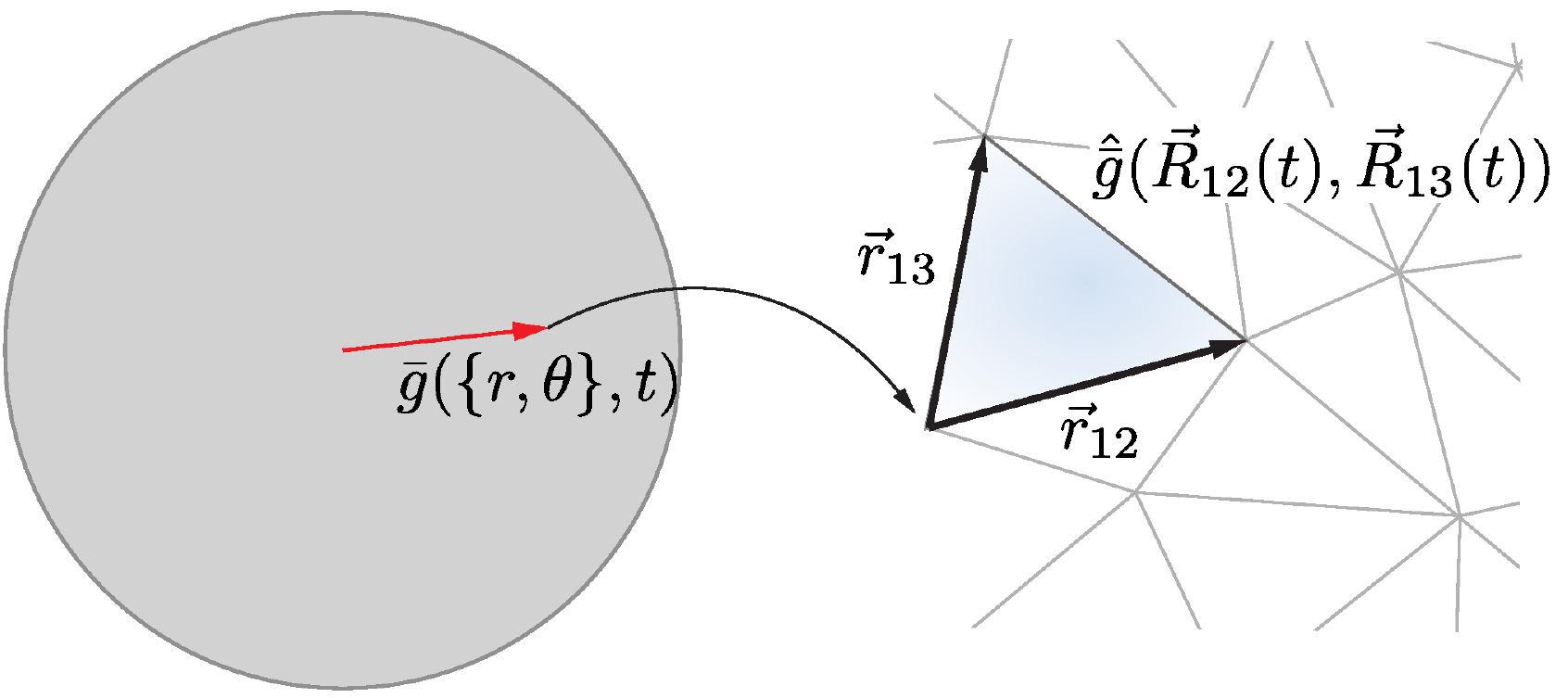}
      \par\end{centering}
      \caption{Active remodelling is introduced as a change of the reference metric of each triangle.\label{fig:growth}}
      
  \end{figure}

  Remodelling is introduced as a change in the local reference metric
  $\bar{g}$. Here we choose a circular geometry for which we have the
  following natural metric (Fig.\ \ref{fig:growth}),
  \begin{equation}
  \bar{g}_{ij}(\{r,\theta\},t)=\left(\begin{array}{cc}
  g_{11}(\{r,\theta\},t) & g_{12}(\{r,\theta\},t)\\
  g_{12}(\{r,\theta\},t) & g_{22}(\{r,\theta\},t)
  \end{array}\right),\label{eq:radial_growth}
  \end{equation}
  and for simplicity we impose a non-shear linear uniform remodelling,
  i.e.,
  \begin{equation}\begin{split}
  \partial_{t}g_{11}(\{r,\theta\},t) & =\beta_{11}\\
  \partial_{t}g_{12}(\{r,\theta\},t) & =\frac{1}{2}\left[\left(\frac{g_{22}}{g_{11}}\right)^{\frac{1}{2}}\beta_{11}+\left(\frac{g_{11}}{g_{22}}\right)^{\frac{1}{2}}\beta_{22}\right]\cos\phi\label{eq:radial_linear_growth}\\
  \partial_{t}g_{22}(\{r,\theta\},t) & =\beta_{22},
  \end{split}\end{equation}
  with $\beta_{11}$ and $\beta_{22}$ being the remodelling rates in the
  $\mathbf{R}_{12}$ and $\mathbf{R}_{13}$ direction and $\phi$ is the angle
  between $\mathbf{R}_{12}$ and $\mathbf{R}_{13}$. Eqs.\ (\ref{eq:radial_growth})
  and (\ref{eq:radial_linear_growth}) can be easily discretised by
  expressing the metric tensor in the triangle laboratory coordinates. 
  
  \section{Viscoelastic relaxation}
  
  As in Ref.\ \cite{Muoz} we model viscoelastic dissipation via an internal rearrangement processes leading to the relaxation of the reference metric towards the
  realised (i.e., current) metric, i.e., 
  \begin{equation}
  \label{eq:viscous-remodelling}
  \partial_t{\bar{g}_{ij}} = \frac{1}{\tau_{ve}}\left(g_{ij} - \bar{g}_{ij}\right),
  \end{equation}
  where ${g}_{ij}$, $\bar{g}_{ij}$ are the current and reference metric tensors at time $t$, respectively, and $\tau_{ve}$ is the remodelling time scale. 
  The physical interpretation of Eq.\ (\ref{eq:viscous-remodelling}) is that the energy is dissipated in local rearrangement 
  processes.
  
  \section{Numerical implementation}
  
\begin{figure*}
  \begin{centering}
  \includegraphics[scale=0.5]{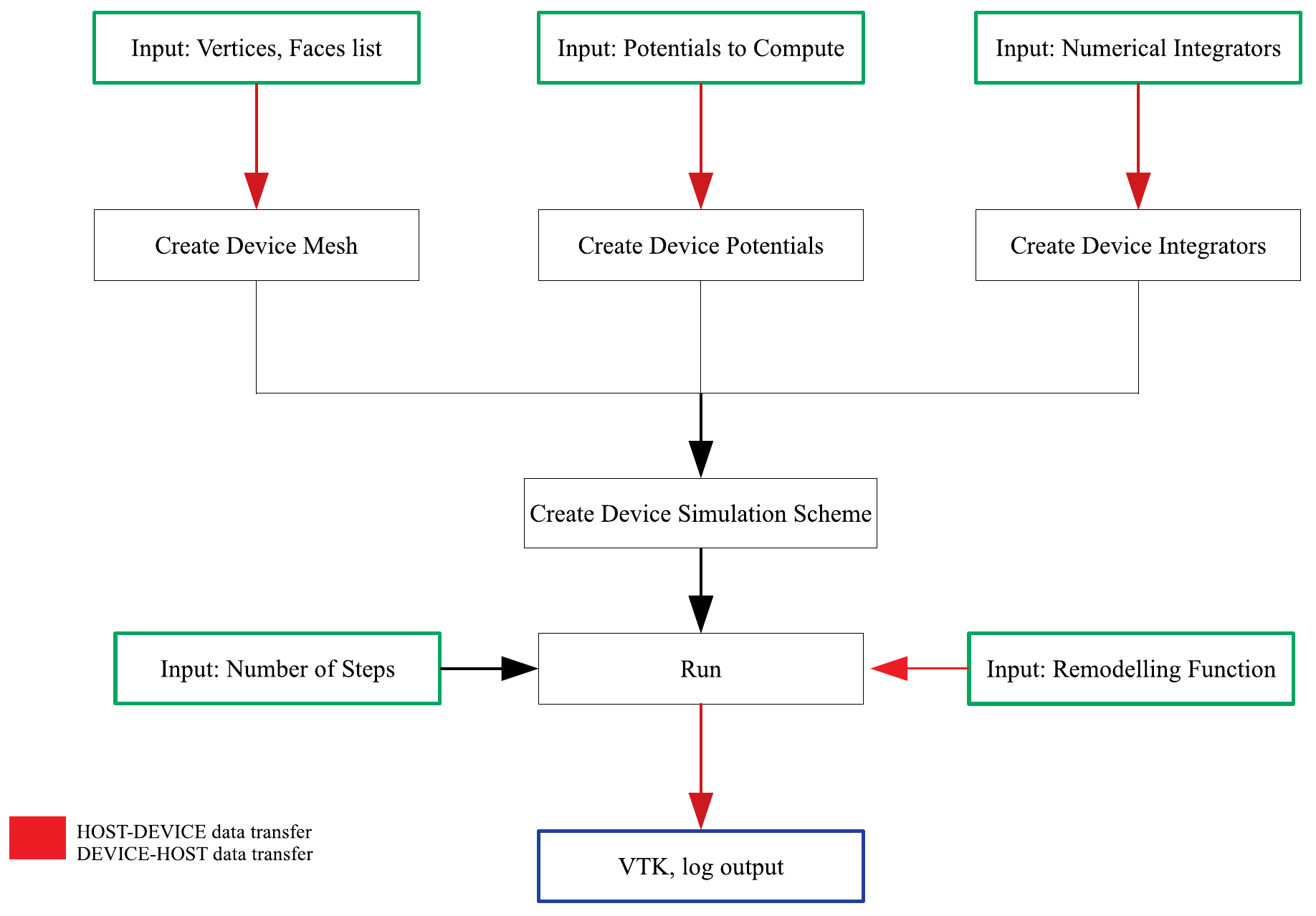}
  \par\end{centering}
  \caption{Schematic diagram of the software implementation. \label{fig:Simulation-Algorithm}}
  
  \end{figure*}

  We have built our own parallel
  GPU-based (NVidia CUDA) implementation of the discrete model outlined in the previous sections. Our code is specifically designed
  to introduce different sources of activity into the system. The general workflow is shown
  in Fig.\ \ref{fig:Simulation-Algorithm}. All computation-heavy task are fully
  implemented on the GPU, so that there are no transfers between DEVICE-HOST
  during the execution. The only routines executed by the host are
  those required by the user in order to save data. 
  
  Our CUDA kernels are moderately optimized, trying to keep aligned
  and coalesced memory access avoiding threads divergence and only using
  atomic functions when absolutely necessary. Finally, we used ParaView \cite{paraview} as an external visualisation software for
  testing and presentation purposes.
  
  \subsection{Simulation parameters}
  
  The coarse-grained triangular meshes used in the simulation were
  created using a public domain package Gmsh \citep{gmesh} setting
  the edge target length to $l=0.35$ and the plate radius equal to
  $R=50$, with all lengths measured in units of thickness, $h$. 
  In order to obtain different initial configuration the vertices
  are moved randomly in $(x,y)$ around the initial configuration using
  a normal distribution with standard deviation equal to $10^{-3}$.
  After this procedure the Device Mesh is created and the reference
  metric is set to the mesh actual metric.
  
  The potentials used in our simulation with its respective parameters
  are listed in Table \ref{tab:Simulation-parameters}. It is important
  to note that all material parameter are assumed to be time-independent and uniform across
  the entire mesh.
  
  \begin{table}[h!]
  \begin{centering}
  \begin{tabular}{|l|c|}
  \hline 
  \textbf{Streaching Potential} & Value\tabularnewline
  \hline 
  \hline 
  Young's modulus, $E$ & $10^{2}$\tabularnewline
  \hline 
  Plate thickness, $h$ & $10^{0}$\tabularnewline
  \hline 
  Poissson's ratio, $\nu$ & $1/3$\tabularnewline
  \hline 
  \textbf{Seung-Nelson Bending Potential} & \tabularnewline
  \hline 
  Bending modulus, $\kappa$ & $5\times10^{-2}(E\,h^3)$\tabularnewline
  \hline 
  \end{tabular}
  \par\end{centering}
  \caption{Simulation parameters.\label{tab:Simulation-parameters}}
  
  \end{table}
  The active remodelling processes are assumed not to be uniform
  on the mesh. In particular, we have chosen to restrict remodelling and
  remodelling to an external annulus of $20<r<50$. The remodelling
  and viscous remodelling rate are set to be uniform inside of the annulus,
  for the respective values used in the simulation, see Fig.\ 3. 
  
  To integrate the vertex equation of motion, we have implemented a Brownian dynamics integrator,
  \begin{equation*}
  \partial_{t}\mathbf{r}_{i}=\mu\mathbf{F}_{i}+\mathbf{F}_{R},
  \end{equation*}
  where $\mu$ is the inverse friction coefficient and $\mathbf{F}_{i}$
  is the total force acting on the vertex $i$ due the mesh deformation
  and $\mathbf{F}_R$ is a uniform random force whose magnitude fullfil
  the fluctuation-dissipation theorem for the given inverse friction
  coefficient and temperature, $T$; in our simulation we set $\mu=1.0$
  and $T=10^{-6}$. In addition, the integration is set to be $10^{-3}$
  for remodelling rates equal or smaller than $10^{-3}$ and $10^{-5}$
  otherwise. 
  
  \section{Elastic relaxation time}
  
  Here we make a rough estimate of the elastic relaxation time scale assuming that the a nearly flat 
  sheet is suspended in a fluid. We assume that the fluid only provides drag and do not consider any 
  effects of its flow, i.e., the fluid acts as a simple sink for the sheet's momentum. Assuming only out of plane 
  motion described by the high function $w(x,y)$, in the overdamped limit the equation of motion is,
  \begin{equation*}
      \Gamma \partial_t w = \kappa \Delta^2 w,
  \end{equation*} 
  where $\Delta^2$ is the bi-laplacian operator and $\Gamma$ is the friction coefficient due to fluid. Note that $\Gamma$ has dimensions of 
  $\frac{\text{mass}}{\text{length}^2\times\text{time}}$, and is thus interpreted as the friction per 
  unit area. If we recall the well-known result in fluid dynamics \citep{Lamb} that the drag coefficient on 
  a disk of radius $R$ moving perpendicular to its plane in a fluid of viscoelastic $\eta$ is $\zeta = 16\eta R v$, where 
  $v$ is the velocity, we obtain $\Gamma = \frac{16}{\pi}\frac{\eta v}{R}$. Therefore, we estimate
  \begin{equation*}
      \Gamma \frac{h}{\tau_{el}} = \kappa \frac{h}{R^4},
  \end{equation*} 
  or
  \begin{equation*}
      \tau_{el} = \frac{16}{\pi}\frac{\eta R^3}{\kappa}.
  \end{equation*}
  For an epithelial tissue of size $R\sim 1$ mm in water, assuming bending rigidity $\kappa\sim10^{-12}$ J, we estimate $\tau_{el}\sim 10$ s.

\end{document}